\makeatletter
\def\input@path{{"C:/Arbeiten/Fertige Arbeiten, Arxive/"}}
\makeatother
\documentclass[english]{article}
\usepackage[T1]{fontenc}
\usepackage[utf8x]{inputenc}
\usepackage{amsmath}
\usepackage{amssymb}
\usepackage{babel}
\begin{document}
\title{Junction Conditions and local Spacetimes in General Relativity}
\author{Albert Huber\thanks{hubera@technikum-wien.at}}
\date{{\footnotesize{}UAS Technikum Wien - Department Applied Mathematics
and Physics, Höchstädtplatz 6, 1200 Vienna, Austria}}
\maketitle
\begin{abstract}
In the present work, a theoretical framework focussing on local geometric
deformations is introduced in order to cope with the problem of how
to join spacetimes with different geometries and physical properties.
This framework is used to show that two Lorentzian manifolds can be
matched by considering local deformations of the associated spacetime
metrics. Based on the fact that metrics can be suitably matched in
this way, it is shown that the underlying geometric approach allows
the characterization of local spacetimes in General Relativity. Furthermore,
it is shown that said approach not only extends conventional thin-shell
formalism, but also allows the treatment of geometric problems that
cannot be treated with standard gluing techniques. 
\end{abstract}
\textit{\footnotesize{}Key words: general relativity, local spacetimes,
junction conditions}{\footnotesize\par}

\section*{Introduction}

The General Theory of Relativity, like most of its countless generalizations,
is a nonlinear theory of gravity. For this reason, it allows the existence
of different types of solutions of Einstein's field equations and
thus the coexistence and co-evolution of different types of gravitational
fields.

However, the fact that the theory allows for a variety of different
solutions, among which many lead to predictions that are in complete
agreement with observation, leads to the challenging mathematical
problem of how pairs of geometrically distinct spacetimes can be joined
with each other along a hypersurface separating the corresponding
Lorentzian manifolds and thus combined into a single geometric field.
This very problem is usually addressed in General Relativity by considering
local junction conditions, the fulfilment of which ensures that the
intrinsic geometric properties of spacetime at the boundary are such
that the local geometries fit together and can therefore be joined
with each other. In this respect, two cases have to be distinguished: 

On the one hand, there is the case in which the boundary hypersurface
is either space- or timelike. In this case, the well-established Darmois-Israel
formalism can be used, which requires the first and second fundamental
forms of given spacetimes to match with each other across the space-
or timelike boundary hypersurface \cite{darmois1927equations,israel1966singular}.
This matching must be given in such a way that the first fundamental
forms of the corresponding geometric fields are continuous and coincide
at the boundary, whereas the second fundamental forms do not have
to coincide and are allowed to be discontinuous across the hypersurface
separating the Lorentzian manifolds \cite{clarke1987junction,israel1966singular,mars1993geometry}.
The condition for this to be the case, however, is the existence of
a concentrated, singular matter distribution - a so-called thin shell
of matter - that happens to form a joint boundary layer for both spacetimes.

On the the hand, there is the case in which the boundary is lightlike
\cite{barrabes1991thin,mars1993geometry}, which is more sophisticated
from a geometrical point of view. In this case, different types of
junction conditions have to be considered, which allow the pairwise
identification of (projections of) gradients of the corresponding
null vector fields across the boundary. The method used then makes
it possible to glue together spacetimes that are separated by a lightlike
boundary hyperface. However, this is not the only advantage offered
by that approach: As it turns out, the geometric framework used is
more general than the Darmois-Israel framework, since it allows one
to combine the null and non-null formalism into a single formalism
called general thin-shell formalism, and also to formulate associated
junction conditions that are always valid regardless of the causal
structure of the boundary hyperface \cite{mars1993geometry,senovilla2015double,senovilla2018equations}.

The main problem that arises in this context, however, is that it
often proves difficult to actually meet said junction conditions,
especially for spacetimes with different causal structures and symmetry
properties. For this reason, it is a rather common case that spacetime
pairs with excessively different geometrical properties cannot be
glued together. This is, moreover, also one of the main reasons why
the above-mentioned conditions have so far been successfully applied
mainly to spacetimes with a comparatively high degree of symmetry,
i.e. to spacetimes with either spherical, cylindrical or plane symmetry,
while the treatment of the problem of how to glue less symmetric spacetimes,
such as stationary, axisymmetric or even non-stationary ones, has
so far received far less attention in the literature.

A primary cause of this shortcoming is the fact that the methods being
used to cope with the junction conditions, in spite of leading to
appropriate local discontinuities in the curvature of respective gravitational
fields (at most 'delta-like'-singularities), in different cases, fail
to deliver physically feasible predictions. The usual reason for this
drawback is the fact that the corresponding methods lead to concentrated,
singular gravitational source terms that often do not obey the energy
conditions of the theory, which significantly reduces their phenomenological
and physical relevance. An even greater problem occurs, moreover,
when the two spacetimes to be joined have metrics of low regularity;
a case that may lead to undefined products of distributions when calculating
the associated curvature fields, such as in the case of gravitational
shock wave spacetimes, where said fields contain 'squares' of the
delta distribution. In this case, the thin-shell formalism obviously
suffers from severe mathematical problems.

In conclusion, however, there appears to be a need for a more reliable
geometric framework that does not lead to unphysical gravitational
source terms and, in contrast to traditional spacetime gluing approaches,
also allows the gluing of distributional metrics that contain terms
proportional to Dirac's delta distribution.

In response to that fact, the aim of a major part of this work is
to provide a simple geometric framework that endeavors to avoid the
aforementioned technical and conceptual difficulties, while at the
same time guaranteeing that the junction conditions of the theory
are met. 

The key difference between the model to be developed and former approaches
to the subject is the fact that said model considers the transition
of one spacetime to another as a dynamical deformation process. This
idea is formally realized by deforming the metric of a given background
geometry and showing that the effect of the deformation completely
subsides if appropriate boundary conditions, which follow from the
addressed junction conditions, are imposed. 

Taking advantage of the fact that the geometric structure of any spacetime
metric can be arbitrarily modified by specifying a suitable deformation
term, and that it is also possible to confine oneself only to those
deformations that have compact supports in an embedded subregion of
a given Lorentzian manifold (or go to zero in a suitable limit), the
geometric framework to be presented ensures that the junction conditions
of the theory are met, thereby allowing a rigorous characterization
of local spacetime geometries in General Relativity. 

This will be demonstrated by some specific geometric examples in section
three of this work, which cover spacetimes with low regularity, such
as gravitational shock wave spacetimes, whose curvature cannot be
calculated using standard gluing techniques. In addition, examples
of non-distributional models are given, which likewise cannot be treated
with the standard technical machinary of the general thin shell formalism.
Additionally, examples of spacetime gluings are discussed in order
to illustrate that said formalism occurs as a special case of the
geometric framework presented in this work.

\section{Junction Conditions and Gluings of Spacetimes}

In Einstein's General Theory of Relativity, the situation quite often
occurs that two spacetime partitions $(\mathcal{M}^{\pm},g^{\pm})$
with two associated Lorentzian manifolds $\mathcal{M}^{\pm}=M^{\pm}\cup\partial M^{\pm}$
are given, which are bounded by a hypersurface $\Sigma$ that forms
a part of the boundary of both spacetimes, so that $\Sigma\subset\partial M^{\pm}$
applies. Given this situation, the question arises as to whether or
not both spacetimes can be 'combined' into an ambient spacetime $(\mathcal{M},g)$,
whose manifold is the union of the manifolds of the individual parts
such that $\mathcal{M}\equiv\mathcal{M}^{-}\cup\mathcal{M}^{+}$. 

A relatively straightforward method that allows one to deal with this
question and thus solve the underlying geometric problem is the method
of gluing spacetimes together across a boundary hypersurface $\Sigma\equiv\partial M^{+}\cap\partial M^{-}$,
using the so-called thin shell formalism \cite{barrabes1991thin,clarke1987junction,israel1966singular,mars1993geometry,reina2016junction,senovilla2015double,senovilla2018equations}. 

According to this method, which has a rich history and important applications
in General Relativity, it is usually assumed that an ambient spacetime
$(\mathcal{M},g)$ with above-mentioned properties is given, i.e.
a spacetime with Lorentzian manifold $M=M^{+}\cup\Sigma\cup M^{-}$
and metric $g_{ab}$, which reduces to the metrics $g_{ab}^{\pm}$
in $M^{\pm}$. As a basis for this, it is further assumed that there
is a restricted $C^{2}$-metric $g_{ab}^{+}=g_{ab}\vert_{M^{+}}$
associated with the part $(\mathcal{M}^{+},g^{+})$ and another $C^{2}$-metric
$g_{ab}^{-}=g_{ab}\vert_{M^{-}}$ associated with $(\mathcal{M}^{-},g^{-})$,
respectively; parts, in relation to which the metric $g_{ab}$ of
the ambient spacetime $(\mathcal{M},g)$ can be decomposed in the
form
\begin{equation}
g_{ab}=\theta g_{ab}^{+}+(1-\theta)g_{ab}^{-},
\end{equation}
where $\theta$ is the Heavyside step function. This step function
is usually assumed to take a value of one half for points lying on
$\Sigma$, a value of one for points lying in $M^{+}$ and a value
of zero for points lying in $M^{-}$. This makes sense as long as
it is ensured that the ambient metric is continuous across the layer,
which implies that (in appropriate coordinates) it must apply that

\begin{equation}
[g_{ab}]=0,
\end{equation}
where $[g_{ab}]\equiv\underset{x\overset{M^{+}}{\rightarrow}x_{0}}{\lim}g_{ab}^{+}(x)-\underset{x\overset{M^{-}}{\rightarrow}x_{0}}{\lim}g_{ab}^{-}(x)$
applies for all $x_{0}\in\Sigma$. 

To provide a coordinate independent description in this context, it
is only natural to use a formalism that is compatible with the intrinsic
geometric structure of the boundary portion $\Sigma$. However, since
such a description cannot be independent of the causal structure of
$\Sigma$, it seems convenient to first closely pursue the most essential
ideas of the so-called general (or mixed) thin shell formalism developed
in \cite{mars1993geometry}, not least because said formalism, unlike
previous approaches to the subject, allows the treatment of the current
physical problem of joining pairs of spacetimes with different geometries
without having to fix the geometric character of the boundary portion
$\Sigma$. Rather, $\Sigma$ can very well be null somewhere in spacetime
and non-null elsewhere.

To enable such treatment of the problem, the formalism takes advantage
of the fact that a pair of normal vector fields $\zeta_{\pm}^{a}$
exists on each side of the layer $\Sigma$ such that $\zeta_{+}^{a}$
corresponds to $M^{+}$ and $\zeta_{-}^{a}$ corresponds to $M^{-}$.
In addition, the fact is exploited that - regardless of the causal
structure of the boundary portion - bases of vector fields $\{E_{\:\rho}^{a}\}$
can be chosen in $T(\Sigma)$ with $\rho=1,2,3$ as well as associated
co-bases $\{e_{\:a}^{\rho}\}$ in $T^{*}(\Sigma)$ such that $e_{\:a}^{\rho}E_{\:\sigma}^{a}=\delta_{\:\sigma}^{\rho}$,
$\zeta_{\pm}^{a}e_{\:a}^{\rho}=0$ and $g_{ab}^{\pm}\vert_{\Sigma}E_{\:\rho}^{a}E_{\:\sigma}^{a}=g_{ab}^{\mp}\vert_{\Sigma}E_{\:\rho}^{a}E_{\:\sigma}^{a}$.
Furthermore, it is observed that there is a pair of vector fields
$\xi_{\pm}^{a}$, usually called rigging vector fields, and an associated
pair of co-vector fields $\xi_{a}^{\pm}$ such that $\xi_{a}^{\pm}\zeta_{\pm}^{a}=-1$
and $\xi_{a}^{\pm}E_{\:\rho}^{a}=0$. The corresponding rigging vector
fields are fixed in this context by demanding $g_{ab}\vert_{\Sigma}\xi_{+}^{a}E_{\:\rho}^{b}=g_{ab}\vert_{\Sigma}\xi_{-}^{a}E_{\:\rho}^{b}\equiv g_{ab}\vert_{\Sigma}\xi^{a}E_{\:\rho}^{b}$
and $g_{ab}\vert_{\Sigma}\xi_{+}^{a}\xi_{+}^{a}=g_{ab}\vert_{\Sigma}\xi_{-}^{a}\xi_{-}^{a}$,
so that the two bases on the tangent spaces $\{\xi_{\pm}^{a},E_{\:\rho}^{a}\}\equiv\{\xi^{a},E_{\:\rho}^{a}\}$
are identified and the $(\pm)$ can be dropped. The two one-forms
$\zeta_{a}^{\pm}$ are automatically identified as well, so that $\{\zeta_{a}^{\pm},e_{a}^{\rho}\}\equiv\{\zeta_{a},e_{a}^{\rho}\}$.
Consequently, it then turns out to be possible to construct a projector
of the type $o_{\,a}^{c}=\delta{}_{\,a}^{c}+\zeta^{c}\xi_{a}$ with
the properties $o_{\,a}^{c}o_{\,b}^{a}=o_{\,b}^{c}$ and $o_{\,a}^{c}\xi^{a}=o_{\,a}^{c}\zeta_{c}=0$,
which can be used as a projector onto $\Sigma$.

With these definitions at hand, the difference (or jump) of any object
from the $+$ or the $-$ sides of $\Sigma$ can be specified. In
particular, any $(m,n)$-tensor field with definite limits on $\Sigma$
from $M^{\pm}$ (regardless of whether it is discontinuous across
$\Sigma$ or not) can be split up in a $+$-part and a $-$-part ,
so that
\begin{equation}
T_{\quad b_{1}b_{2}...b_{n}}^{a_{1}a_{2}...a_{m}}=\theta T_{\quad b_{1}b_{2}...b_{n}}^{+a_{1}a_{2}...a_{m}}+(1-\theta)T_{\quad b_{1}b_{2}...b_{n}}^{-a_{1}a_{2}...a_{m}}.
\end{equation}
The covariant derivative of the same object then reads
\begin{equation}
\nabla_{c}T_{\quad b_{1}b_{2}...b_{n}}^{a_{1}a_{2}...a_{m}}=\theta\nabla_{c}^{+}T_{\quad b_{1}b_{2}...b_{n}}^{+a_{1}a_{2}...a_{m}}+(1-\theta)\nabla_{c}^{-}T_{\quad b_{1}b_{2}...b_{n}}^{-a_{1}a_{2}...a_{m}}+\delta_{c}[T_{\quad b_{1}b_{2}...b_{n}}^{a_{1}a_{2}...a_{m}}],
\end{equation}
where $[T_{\quad b_{1}b_{2}...b_{n}}^{a_{1}a_{2}...a_{m}}]\equiv\underset{x\overset{M^{+}}{\rightarrow}x_{0}}{\lim}T_{\quad b_{1}b_{2}...b_{n}}^{+a_{1}a_{2}...a_{m}}(x)-\underset{x\overset{M^{-}}{\rightarrow}x_{0}}{\lim}T_{\quad b_{1}b_{2}...b_{n}}^{-a_{1}a_{2}...a_{m}}(x)$
applies for all $x_{0}\in\Sigma$ and $\delta_{c}\equiv\zeta_{c}\delta$
is a vector-valued distribution constructed from Dirac's delta distribution
$\delta=\delta(x)$.

As already mentioned above, given a suitable pair of coordinate charts
$(x_{\pm}^{0},x_{\pm}^{1},x_{\pm}^{2},x_{\pm}^{3})$, the metric is
continuous across $\Sigma$. However, its derivatives, and thus the
corresponding connections, are discontinuous. In fact, it is found
in this context that

\begin{equation}
[\partial_{c}g_{ab}]=2\cdot\zeta_{c}\gamma_{ab}
\end{equation}
and therefore 
\begin{equation}
[\Gamma_{\,bc}^{a}]=\gamma_{b}^{\,a}\zeta_{c}+\gamma_{c}^{\,a}\zeta_{b}-\gamma_{bc}\zeta^{a},
\end{equation}
where $\gamma_{ab}$ is a symmetric tensor field defining the properties
of the shell.

The associated Riemann tensor is of the form

\begin{equation}
R_{\,bcd}^{a}=\theta R_{\,bcd}^{+a}+(1-\theta)R_{\,bcd}^{-a}+\delta H_{\,bcd}^{a},
\end{equation}
where $\delta$ is the Dirac delta distribution and $H_{\,bcd}^{a}$
represents the singular part of the curvature tensor distribution,
which is explicitly given by $H_{\,bcd}^{a}=\frac{1}{2}(\gamma_{\,d}^{a}\zeta_{b}\zeta_{c}-\gamma_{\,c}^{a}\zeta_{b}\zeta_{d}+\gamma{}_{bc}\zeta^{a}\zeta_{d}-\gamma_{bd}\zeta^{a}\zeta_{c}).$ 

Given this definition, the said approach allows for a generalized
formulation of Einstein's field equations in a distributional sense,
which leads to a distributional Einstein tensor of the form

\begin{equation}
G_{\,b}^{a}=\theta G_{\,b}^{+a}+(1-\theta)G_{\,b}^{-a}+\delta\cdot\rho_{\;b}^{a},
\end{equation}
where $\rho_{\;b}^{a}=H_{\;b}^{a}-\frac{1}{2}\delta_{\;b}^{a}H$ (with
$H\equiv g^{ab}\vert_{\Sigma}H_{ab}$) is a symmetric covariant tensor
field defined only at points of the hypersurface $\Sigma$. The associated
stress-energy tensor, on the other hand, has to possess the form

\begin{equation}
T_{\,b}^{a}=\theta T_{\;b}^{+a}+(1-\theta)T_{\;b}^{-a}+\delta\cdot\tau_{\;b}^{a},
\end{equation}
which is why, of course, it is required that $G_{\,b}^{\pm a}=8\pi T_{\;b}^{\pm a}$
and $\rho_{\;b}^{a}=8\pi\tau_{\;b}^{a}$. 

On the basis of these relations (and some others which will not be
specifically relevant for this work), the underlying thin-shell formalism
can be used to join different partitions of spacetime, including those
where parts of $\Sigma$ are either null or non-null, that is, either
null or spacelike or timelike. 

In the latter cases, where portions of the thin shell are allowed
to be non-null, the above general formalism traces back to the Darmois-Israel
method, which is based on a $3+1$-decomposition of spacetime. The
geometric setting used in this method is essentially the same as that
of the general formalism mentioned above, with the only exception
being that it is additionally required a priori that the ambient spacetime
$(\mathcal{M},g)$ admits a foliation in either spacelike or timelike
hypersurfaces and a boundary portion $\Sigma$ (with fixed causal
structure) that can be embedded in said foliation of spacetime. This
additional restriction of the geometry of spacetime allows the consideration
of a congruence of curves generated by a normalized vector field $n^{a}$
that is necessarily orthogonal to $\Sigma$ and therefore fulfills
$n_{a}n^{a}=\epsilon$ and $\nabla_{[c}n_{a]}=0$ with $\epsilon=\pm1$.
This normal vector field (and its associated co-normal) can then be
used to define the first and second fundamental forms $h_{ab}=g_{ab}+\epsilon n_{a}n_{b}$
and $K_{ab}=h_{\,a}^{c}h_{\,b}^{d}\nabla_{(c}n_{d)}$. 

In order to find the corresponding shell equations and to guarantee
that the spacetime partitions $(\mathcal{M}^{\pm},g^{\pm})$ can be
'combined' to the ambient spacetime $(\mathcal{M},g)$ in the given
case, however, it must be ensured that the said pair of spacetimes
exhibits spacelike or timelike foliations compatible to that of the
ambient spacetime $(\mathcal{M},g)$. Essentially, this means that
pairs of either timelike or spacelike generating vector fields $n_{\pm}^{a}$
with the properties $n_{a}^{\pm}n_{\pm}^{a}=\epsilon$ and $\nabla_{[c}^{\pm}n_{a]}^{\pm}=0$
must exist, which can be appropriately identified across the shell
in a manner similar to the general formalism. By definition, these
vector fields have to be orthogonal to the respective first and second
fundamental forms $h_{ab}^{\pm}=g_{ab}^{\pm}+\epsilon n_{a}^{\pm}n_{b}^{\pm}$
and $K_{ab}^{\pm}=h_{\,a}^{\pm c}h_{\,b}^{\pm d}\nabla_{(c}^{\pm}n_{d)}^{\pm}$.
In addition, it needs to be assumed that the three-metrics of the
spacetime partitions are, at least, continuous across $\Sigma$, so
that $[h_{ab}]=0$ is valid; although $[K_{ab}]=0$ does not necessarily
have to apply in this context. The corresponding shell equations then
yield conditions for matching the Cauchy data $(h_{ab}^{\pm},K_{ab}^{\pm})$
of the bounded spacetimes $(\mathcal{M}^{\pm},g^{\pm})$ in such a
way across $\Sigma$ that they are consistent with the Cauchy data
$(h_{ab},K_{ab})$ of the ambient spacetime $(\mathcal{M},g)$. These
shell equations result directly from the general formalism discussed
above if one sets $\xi_{a}\equiv\epsilon n_{a}$, $\zeta^{a}\equiv n^{a}$,
$o_{\;a}^{c}\equiv h_{\;a}^{c}=\delta_{\;a}^{c}+\epsilon n^{c}n_{a}$
and $\gamma{}_{ab}=\epsilon[K_{ab}]$ in relations $(5-9)$, thus
proving the fact that said formalism actually contains the Darmois-Israel
framework as a special case.

Accordingly, in order to avoid the existence of ill-defined singular
contributions to the field equations, it must be required that the
pairs of first and the second fundamental forms associated with pairs
of spacetimes $(\mathcal{M}^{\pm},g^{\pm})$ satisfy the junction
conditions 

\begin{equation}
[h_{ab}]=0
\end{equation}
and

\begin{equation}
[K_{ab}]=8\pi\epsilon(\tau_{ab}-\frac{1}{2}h_{ab}\tau);
\end{equation}
conditions which represent, in a quite generic way, geometrically
necessary requirements for the identification of the corresponding
Lorentzian manifolds.

It is worth noting that even though the given approach is formulated
in a coordinate-independent manner, it still leads back to alternative
formulations of junction conditions, for example those given by Lichnerowicz
or O'Brian and Synge in case of the special choice of so-called admissible
coordinates on both sides of the layer \cite{israel1966singular}. 

Furthermore, there is also the case that $\Sigma$ is locally null;
a case that seems to require a more deliberate approach than the traditional
non-null description of the problem not least due to the fact that
the first fundamental form is degenerate on a null hypersurface and
the associated null normal is not only orthogonal but also tangential
to it. Nevertheless, assuming the existence of two pairs of null congruences
generated by a pair of lightlike vector fields $l^{a}$ and $k^{a}$
that are orthogonal to a spacelike two-slice $S\subset\Sigma$, it
turns out that the general formalism is versatile enough to include
said special case as well; giving rise to the same shell equations
previously found in \cite{barrabes1991thin}. These equations can
be obtained from expressions $(5-9)$ if the choice $\xi_{a}\equiv l_{a}$,
$\zeta^{a}\equiv k^{a}$ and $o_{\,a}^{c}=\delta{}_{\,a}^{c}+k^{c}l_{a}$
is made in this context and a covariant symmetric two-form $\mathcal{H}_{ab}=2o_{\,a}^{c}o_{\,b}^{d}\nabla_{(c}l_{d)}$
is specified, which (in the continuous null limit of the Darmois-Israel
framework) has the same jump discontinuity features as the extrinsic
curvature across $\Sigma$. More precisely, while in the non-null
case generically one has $\gamma_{bc}\equiv\epsilon[K_{bc}]$, in
the lightlike case, where there is a different geometric setting,
one has $\gamma_{ab}\equiv[\mathcal{H}_{ab}]$. With regard to this
specific quantity, which shall be called Mars-Senovilla two-form from
now on, the above series of junction relations turns into

\begin{equation}
[\mathcal{H}_{ab}]=0
\end{equation}
and therefore becomes, in contrast to the non-null case in which the
right hand side is non-vanishing, a trivial set of relations; relations
that guarantee that the Einstein tensor of the geometry contains no
singular part proportional to Dirac's delta distribution (and thus
no surface layer).

However, as shown in \cite{mars1993geometry}, these junction relations
can actually be relaxed in the given null case by requiring that the
respective Mars-Senovilla two-forms meet the conditions
\begin{equation}
[\mathcal{H}_{ab}]k^{b}=[\mathcal{H}]=0,
\end{equation}
which also guarantee that the singular part of the curvature tensor
distribution vanishes identically.

Anyhow, junction conditions do not necessarily have to be based on
a $3+1$-decomposition of spacetime; they also have been formulated
in the $2+2$-framework toward General Relativity or in space-time
approaches that are based on a $1+1+2$-decomposition of spacetime.

In particular, as shown by Penrose in \cite{penrose1972general},
junction conditions can be formulated (in a coordinate dependent manner)
which are based on a dual null foliation of spacetime in spacelike
surfaces. These conditions form the basis of Penrose's now infamous
cut-and-paste method, which provides the formal basis for the description
of gravitational shock wave spacetimes in General Relativity and turns
out to be closely related to the thin shell framework in specific
applications. 

Besides that, in order to characterize boundary portions that possess
a 'corner' or a 'sharp edge', another set of junction relations has
been formulated in the literature in the past \cite{hayward1993gravitational}.
For the purpose of formulating said conditions, a timelike generating
vector field $n^{a}$ and a spacelike one $u^{a}$ associated with
respective timelike and spacelike congruences have been considered,
which yield spacelike and timelike foliations and thus a $1+1+2$-decomposition
of spacetime. Based on the existence of said foliations, the fact
was exploited that the spacetime metric $g_{ab}$ can be decomposed
in the form $g_{ab}=-n_{a}n_{b}+h_{ab}=u_{a}u_{b}+\gamma_{ab}$ with
$h_{ab}=q_{ab}+s_{a}s_{b}$ and $\gamma_{ab}=q_{ab}-v_{a}v_{b}$,
respectively, where $s^{a}$ and $v^{a}$ are spacelike and timelike
unit normals orthogonal to $n^{a}$ and $u^{a}$ and $q_{ab}$ is
the induced Riemannian metric on a spacelike two-slice $S\subset\Sigma$.
Due to the fact that the given vector fields $n^{a}$ and $u^{a}$
are not assumed to be normalized with respect to each other in this
context, it then typically turns out that one has to deal (in the
case of a spacelike joint) with a non-vanishing edge 'angle' $\Theta=\cos^{-1}((n,u))$
in such approaches. This non-vanishing quantity has been shown to
lead to jump discontinuities and therefore to an additional set of
junction conditions given by

\begin{equation}
[\Theta]q_{ab}=\mathcal{T}_{ab},
\end{equation}
where $\mathcal{T}_{ab}$ is the stress-energy tensor restricted to
$S$. This particular set of conditions completes the list of junction
conditions discussed in this work. In the following, however, only
conditions $(10)$ and $(11)$ or conditions $(12)$ and $(13)$ will
prove to be really relevant, since the validity of these conditions
will be used as a basis for a geometric extension of the general thin
shell formalism. 

The reason why such a geometric extension proves useful (or even necessary)
is the following: When using thin-shell formalism one must expect
that the boundaries of the spacetimes to be glued together can be
singular hyperfaces. But that means that the curvature along these
hypersurfaces can grow infinitely, which is problematic as long as
no plausible physical reason for the occurrence of such infinities
is given, such as possibly the occurrence of a relativistic shock
wave at the boundary of the spacetimes or something similar. Besides
that, the general thin shell formalism suffers from the problem that
the singular parts of the energy-momentum tensor occurring in $(9)$
may fail to obey relevant energy conditions and, what is even worse,
it turns out that said conditions cannot even be formulated in all
cases of relevance, such as, in particular, in the case of the dominant
energy condition. The main reason for this drawback is that in order
to set up the dominant energy condition, one would have to deal ill-defined
products of distributions, that is, with 'squares' of the delta distribution.

The exact same problem occurs if the metrics $g_{ab}^{\pm}$ considered
in $(1)$ are not $C^{2}$-metrics, but have lower regularity. The
situation becomes particularly alarming if one of the metrics of the
two spacetime partitions $(\mathcal{M}^{\pm},g^{\pm})$ contains a
part that is proportional to the Dirac delta distribution. To illustrate
this, the specific case shall be considered in which $g_{ab}^{+}=g_{ab}^{0}+\delta e_{ab}$
and $g_{ab}^{-}=g_{ab}^{0}$, where $g_{ab}^{0}$ is some smooth $C^{2}$-metric
and $e_{ab}$ and is a smooth tensor field. In this case, taking advantage
of the fact $\theta\delta\thickapprox A\delta$, where $A$ is a constant
and $\thickapprox$ means association in the sense of distributions,
splitting $(1)$ yields $g_{ab}\thickapprox g_{ab}^{0}+A\delta e_{ab}$
and equations $(7-9)$ become (similar as in the case of the generic
energy condition) relations that contain 'squares' of the delta distribution.
Consequently, in this particular case not even the field equations
of the theory can be defined meaningfully.

In response to these defeciences of the general thin shell formalism,
the remainder of the present work will address the problem of joining
different spacetimes from a slightly different angle, namely by means
of a geometric approach based on the use of metric deformations. This
approach generalizes the thin shell formalism in such a way that in
important special cases the treatment of the above mentioned problems
becomes possible in full accordance with the junction conditions of
the theory. To make this possible, the following two steps are taken:
First, the junction conditions of the general thin shell formalism
are reformulated in the language of the geometric deformation approach,
and second, the support properties of the corresponding deformation
fields are restricted so that all types of junction conditions discussed
in this section are fulfilled. In this way, as will be explained,
the concept of local spacetime geometry is introcued in relation to
a fixed ambient geometry of spacetime. 

\section{Local Geometries and Deformations of Spacetime}

In order to approach now the subject of joining spacetimes from a
different angle, namely by using special metric deformations that
allow one to meet the junction conditions discussed in the previous
section, two different spacetime partitions $(\mathcal{M}^{\pm},g^{\pm})$
of an ambient spacetime $(\mathcal{M},g)$ shall once more be considered.
These partitions, as before, shall be assumed to be bounded by a hypersurface
$\Sigma$ which forms a part of the boundary of both spacetimes such
that $\Sigma\subset\partial M^{\pm}$ applies. 

Without any further assumptions about the geometric structures of
both of the spacetimes $(\mathcal{M}^{\pm},g^{\pm})$, both partitions
are allowed to exhibit totally different geometric properties anywhere
except for their boundary, where, by the introduced junction conditions,
both spacetimes have to possess identical induced geometries. As a
necessary prerequisite for obtaining a spacetime $(\mathcal{M},g)$
with connected manifold $\mathcal{M}=\mathcal{M}^{+}\cup\mathcal{M}^{-}$,
it must therefore be ensured that said spacetime partitions can be
identified along the boundary portion $\Sigma$ in such a way that
the junction conditions discussed in the previous section are met.

In order to ensure this, the same approach as in the previous section
shall be followed, namely different bases $\{\xi_{\pm}^{a},E_{\:\rho}^{a}\}$
and co-bases $\{\zeta_{a}^{\pm},e_{a}^{\rho}\}$ shall be considered
at each side of the boundary, which can be identified along $\Sigma$
in the same way as in section one of this work.

Against this background, the main task now is to construct a spacetime
$(\mathcal{M},g)$ with connected Lorentzian manifold $\mathcal{M}=\mathcal{M}^{+}\cup\mathcal{M}^{-}$,
with local spacetime partitions $(\mathcal{M}^{\pm},g^{\pm})$ that
may have different geometric structures everywhere except along the
boundary hyperface $\Sigma$. 

To face this task, the following observation proves useful: A change
of one spacetime geometry with respect to another can be characterized
by considering a deformation of the associated metrics, where deformation
in this context means any backreaction that changes the geometric
properties of a given spacetime metric with respect to a given background
metric and a given class or group of deformation fields propagating
on the corresponding background spacetime.

To make this final statement precise, let $g_{ab}$ be a fundamental
metric field associated with the ambient spacetime $(\mathcal{M},g)$
with the manifold structure $\mathcal{M}=\mathcal{M}^{+}\cup\mathcal{M}^{-}$.
Considering the tensor deformations
\begin{equation}
g_{ab}=g_{ab}^{\pm}+e_{ab}^{\pm}
\end{equation}
and
\begin{equation}
g^{ab}=g^{\pm ab}+f^{\pm ab},
\end{equation}
one can define the non-vanishing tensor fields $g_{ab}^{\pm}=g_{ab}-e_{ab}^{\pm}$
and $g^{\pm ab}=g^{ab}-f^{\pm ab}$, which shall be required to be
at least $C^{1}$, although they may actually turn out to be $C_{0}^{\infty}$
(locally) in various cases of interest. These tensor fields a priori
do not represent a metric and an inverse metric, respectively. Instead,
the objects in question are special tensor fields whose properties
depend on the choice of the deformation tensor fields $e_{ab}^{\pm}$
and $f^{\pm ab}$. 

Using these deformation tensor fields, one can re-write various geometric
relations involvong the metric $g_{ab}$ of $(\mathcal{M},g)$ and
its inverse $g^{ab}$. Specifically, provided that that the tensor
fields $e_{a}^{\pm\:b}:=e_{ac}^{\pm}g^{\pm cb}=e_{ac}^{\pm}(g^{cb}-f^{\pm cb})$
and $f_{a}^{\pm\:b}:=g_{ac}^{\pm}f^{\pm cb}=(g_{ab}-e_{ab}^{\pm})f^{\pm cb}$
are identified as tensor fields on $(\mathcal{M},g)$ , the relation

\begin{equation}
g_{ab}g^{bc}=\delta_{a}^{\:c},
\end{equation}
 can be brought into the form 
\begin{equation}
e_{a}^{\pm\:b}+f_{a}^{\pm\:b}+e_{a}^{\pm\:c}f_{c}^{\pm\:b}=0.
\end{equation}
In addition, one can define the difference tensors $C_{\:bc}^{\pm a}=\frac{1}{2}(g^{\pm ad}+f^{\pm ad})(\nabla_{b}^{\pm}e_{dc}^{\pm}+\nabla_{c}^{\pm}e_{bd}^{\pm}-\nabla_{d}^{\pm}e_{bc}^{\pm})$
in relation to the unique Levi-Civita connection defined on $(\mathcal{M},g)$,
which then allows one to decompose the Riemann tensor of the geometry
in the following form

\begin{equation}
R_{\,bcd}^{a}=R_{\,bcd}^{\pm a}+E_{\,bcd}^{\pm a},
\end{equation}
where $E_{\,bcd}^{\pm a}=2\nabla_{[c}^{\pm}C_{\,d]b}^{\pm a}+2C_{e[c}^{\pm a}C_{d]b}^{\pm e}$
shall apply by definition. By contracting indices, one finds that
the Ricci tensor takes the form

\begin{equation}
R_{bd}=R_{\,bd}^{\pm}+E_{bd}^{\pm},
\end{equation}
where $E_{bd}^{\pm}=E_{\,bad}^{\pm a}=E_{bd}^{\pm}=2\nabla_{[a}^{\pm}C_{\,d]b}^{\pm a}+2C_{e[a}^{\pm a}C_{d]b}^{\pm e}$
follows from the foregoing definition. By repeating that procedure,
the decomposition of the Ricci scalar

\begin{equation}
R=R^{\pm}+g^{\pm bd}E_{bd}^{\pm}+f^{\pm bd}R_{bd}^{\pm}+f^{\pm bd}E_{bd}^{\pm}
\end{equation}
can also be obtained. However, as a direct consequence, Einstein's
equations 

\begin{equation}
G_{ab}=8\pi T_{ab}
\end{equation}
can be re-written in the form

\begin{equation}
G_{ab}^{\pm}+\rho_{ab}^{\pm}=8\pi T_{ab},
\end{equation}
provided that $\rho_{ab}^{\pm}=\psi_{ab}^{\pm}-\frac{1}{2}g_{ab}^{\pm}(f^{\pm cd}R_{cd}^{\pm}+f^{\pm cd}E_{cd}^{\pm})-\frac{1}{2}e_{ab}^{\pm}(R^{\pm}+f^{\pm cd}R_{cd}^{\pm}+g^{\pm cd}E_{cd}^{\pm}+f^{\pm cd}E_{cd}^{\pm})$
with $\psi_{ab}^{\pm}=E_{ab}^{\pm}-\frac{1}{2}g_{ab}^{\pm}(g^{\pm cd}E_{cd}^{\pm})$
holds in the given context. 

Having obtained these relations, the following observation can be
made: By requiring that the deformation tensor fields $e_{ab}^{\pm}$
and $f^{\pm ab}$ defined above vanish somewhere in local subregions
of the manifold $\mathcal{M}$, the tensor fields $g_{ab}^{\pm}$
coincide locally with the metric $g_{ab}$ of the spacetime $(\mathcal{M},g)$
. Therefore, the following can be concluded: As long as the tensor
fields $e_{ab}^{\pm}$ and $f^{\pm ab}$ are defined in such a way
that they vanish globally in $\mathcal{M}^{\pm}\subseteq\mathcal{M}$,
which is certainly the case if the components of said fields are $C_{0}^{\infty}$
functions or distributions with compact supports lying in the complements
$\mathcal{M}_{C}^{\pm}\equiv\mathcal{M}\backslash\mathcal{M}^{\pm}$
of the Lorentzian manifolds $\mathcal{M}^{\pm}$, the tensor fields
$g_{ab}^{\pm}$ represent well-defined (possibly even smooth) metric
fields, but only within the local regions $\mathcal{M}^{\pm}$. 'Outside'
these regions, however, they are just specific tensor fields, so that
it can be concluded that the pairs $(\mathcal{M}^{\pm},g^{\pm})$
define pairs of \textit{local spacetimes}, i.e. pairs of spacetimes
whose metrics $g_{ab}^{\pm}$ represent well-defined tensor fields
on $(\mathcal{M},g)$, which coincide locally with the 'correct' metric
of spacetime, which is $g_{ab}$\footnote{In this context, it is important to note that neither $e_{ab}^{\pm}$
nor $f^{\pm ab}$ are assumed to be small and that both pairs of objects
define a whole class of tensor deformations which in principle can
become arbitrarily large, so that for a given vector field $w^{a}$,
which is causal with regard to one of the local metrics $g_{ab}^{\pm}$,
there is no need for it to be causal with regard to the other local
metric $g_{ab}^{\mp}$ or the ambient metric $g_{ab}$ in the complement
$\mathcal{M}\backslash\mathcal{M}^{\pm}$ of $\mathcal{M}^{\pm}$.
However, in the opposite case, the deformations may also become arbitrarily
small, so that they become the subject of relativistic perturbation
theory.}. 

Probably the simplest way to construct deformation fields $e_{ab}^{\pm}$
and $f^{\pm ab}$ with the required properties is to make an ansatz
of the form $e_{ab}^{\pm}=\chi_{\mathcal{M}_{C}^{\pm}}\mathtt{e}_{ab}^{\pm}$,
where the $\chi_{\mathcal{M}_{C}^{\pm}}$ are indicator functions
(also called characteristic functions) with compact support in $\mathcal{M}_{C}^{\pm}\equiv\mathcal{M}\backslash\mathcal{M}^{\pm}$
and $\mathtt{e}_{ab}^{\pm}$ are continuous, at least twice differentiable
tensor fields. Since the Heaviside step function is a special indicator
function, which is at the same time a generalized function, one can
always choose said functions and the corresponding deformation fields
in such a way that the distributional splitting $(1)$ results as
a special case of the given construction. 

However, as it turns out, there is no need in general to require that
$\chi_{\mathcal{M}_{C}^{\pm}}$ are indicator functions. Rather, it
suffices to choose said functions as smooth transition functions (or
a sequence of such functions), which provide a smooth transition from
zero to one in the unit interval $[0,1]$. A transition (alias cut-off)
function with these particular properties can be obtained by considering
the non-analytic smooth function 
\begin{equation}
\psi(x):=\begin{cases}
\overset{e^{-\frac{1}{x}}}{\underset{0}{}} & \overset{x>0}{\underset{x\leq0}{}}\end{cases},
\end{equation}
which meets the conditions $0\leq\psi\leq1$ and $\psi(x)>0$ if and
only if $x>0$. This function can be used to define the transition
function

\begin{equation}
\chi(x)=\frac{\psi(\frac{x_{0}}{x})}{\psi(\frac{x_{0}}{x})-\psi(1-\frac{x_{0}}{x})},
\end{equation}
which contains a constant $x_{0}$ that ensures that the exponent
in $(24)$ is dimensionless and therefore takes a value of zero for
$x<0$, a value of one for $x\geq x_{0}$ and is strictly increasing
in the interval $[0,1]$. This can then be used to give the further
definition

\begin{equation}
1-\chi(x)=\frac{\psi(1-\frac{x_{0}}{x})}{\psi(\frac{x_{0}}{x})-\psi(1-\frac{x_{0}}{x})},
\end{equation}
which has the same properties as $\chi(x)$, but is strictly decreasing.

By using one of these transition functions instead of the Heaviside
step function, a smooth analogon of the distributional splitting $(1)$
can then obtained. It is worth noting that similar approaches can,
of course, be given by considering bump functions or other smooth
functions with similar support properties, which may be constructed
from convolutions of smooth functions with mollifiers. 

Since in all these approaches, the full Einstein equations $(22)$
reduce to the restricted local Einstein equations $G_{ab}^{\pm}=8\pi T_{ab}^{\pm}$
on $(\mathcal{M}^{\pm},g^{\pm})$, it becomes clear that the remaining
equations

\begin{equation}
\rho_{ab}^{\pm}=8\pi\tau_{ab}^{\pm},
\end{equation}
can be determined independently in agreement with the introduced junction
conditions, where, of course, $\tau_{ab}^{\pm}:=T_{ab}-T_{ab}^{\pm}$
applies in the given context.

However, it must be stressed that there is a price to be paid in this
context: By introducing transition functions of the form $(25)$ and
$(26)$, the manifold structure is no longer $\mathcal{M}=\mathcal{M}^{-}\cup\mathcal{M}^{+}$,
but rather $\mathcal{M}=\mathcal{M}^{-}\cup\mathcal{O}\cup\mathcal{M}^{+}$,
where $\mathcal{O}$ is some transition region in which $\chi(x)$
continuously increases until it reaches a value of one. Consequently,
by considering transition function of the above form in order to make
sure that there is a smooth geometric transition between the pairs
of local spacetimes $(\mathcal{M}^{\pm},g^{\pm})$, one ends up in
a situation where has to deal with three spacetime partitions $(\mathcal{M}^{\pm},g^{\pm})$
and $(\mathcal{O},g)$. This implies, however, that one suddenly has
to deal with a completely new geometric setting, which is slightly
different from the one usually considered by general thin-shell formalism.

As a direct consequence, however, the question arises which junction
conditions need to be fulfilled at the boundaries $\Sigma_{\pm}=\partial\text{\ensuremath{\mathcal{M}}}_{\pm}\cap\partial\mathcal{O}$.
Beyond that, generally speaking, the question arises of how the junction
conditions of general thin shell formalism can be formulated to describe
the first case mentioned above and under which circumstances these
conditions can be fulfilled in the given setting. 

To face these questions, one may take a closer look at conditions
$(10-13)$ of the previous section. Sure enough, these conditions,
if they were to be fulfilled, will lead to constraints on the deformation
fields $e_{ab}^{\pm}$ and $f^{\pm ab}$. More specifically, considering
the case in which the manifold structure is $\mathcal{M}=\mathcal{M}^{+}\cup\mathcal{M}^{-}$,
said conditions lead to the requirement that the fields $e_{ab}^{\pm}$
and $f^{\pm ab}$ match exactly at $\Sigma$ and have compact support
in $\mathcal{M}\text{\ensuremath{\backslash\{\mathcal{M}^{\pm}\text{\ensuremath{\backslash\Sigma}}\}}}$,
thereby ensuring that the spacetime partitions $(\mathcal{M}^{\pm},g^{\pm})$
can pointwise be joined along $\Sigma$. 

To see this, one may consider that in the given setting, the junction
conditions $(10)$ and $(11)$ of the Darmois-Israel framework require

\begin{equation}
[e_{ab}]=0
\end{equation}
and
\begin{equation}
n_{a}h_{\,b}^{e}h_{\,c}^{f}[C_{ef}^{a}]=8\pi\epsilon(\tau_{bc}-\frac{1}{2}h_{cb}\tau)
\end{equation}
to hold in a suitable coordinate chart. In the lightlike case, on
the other hand, junction condition $(12)$ requires 
\begin{equation}
l_{a}o_{\,b}^{e}o_{\,c}^{f}[C_{ef}^{a}]=0
\end{equation}
to be fulfilled. Consequently, however, it can be concluded that the
spacetime partitions $(\mathcal{M}^{\pm},g^{\pm})$ can always be
smoothly joined - regardless of the causal structure of the boundary
hyperface - if 

\begin{equation}
[e_{ab}]=[f^{ab}]=0,\:[C_{bc}^{a}]=0
\end{equation}
applies in a suitable coordinate chart. In the smooth case mentioned,
however, the case may very well occur that $(30)$ is valid instead
of $[C_{bc}^{a}]=0$ if $\Sigma$ is null, or that the right side
of $(29)$ is zero if $\Sigma$ is not null, since the condition $[C_{bc}^{a}]=0$
can only be met in special cases. Moreover, since not all spacetimes
can be smoothly joined, it may be required in cases where it is not
possible to fulfill condition $(31)$ that

\begin{equation}
[e_{ab}]=[f^{ab}]=0,\:[C_{bc}^{a}]\neq0
\end{equation}
applies and conditions $(29)$ or $(30)$ are met as well. However,
these are exactly the conditions of the thin shell formalism simply
transferred to the given geometric setting; conditions that are known
to produce reasonable results when the metrics to be glued are $C^{2}$-metrics. 

The reformulation of these conditions in the context of the geometrical
deformation approach developed in this section can be vindicated by
the fact that said approach - in combination with Colombeau's theory
of generalized functions \cite{colombeau2000new,colombeau2011elementary}
- allows an extension of the 'classic' thin-shell formalism. In particular,
as shall be substantiated by concrete examples in the next section,
it becomes possible to glue spacetime metrics that differ by deformation
terms that are proportional to Dirac's delta distribution, but are
nevertheless of such a form that the condition $(32)$ and the conditions
$(29)$ or $(30)$ can still be fulfilled. The reason for this is
that the geometric deformation approach provides direct information
on certain problematic terms and expressions that require careful
treatment or, in other words, need to be studied in more detail using
Colombeau's theory. In this way, the approach enables the treatment
of problems that would go beyond the usual scope of the formalism
due to the low regularity of the spacetime metrics to be glued. Concrete
examples, however, will only be given later - in the next section
of this work.

Anyway, the situation is completely different when the metrics of
the spacetime partitions $(\mathcal{M}^{\pm},g^{\pm})$ are not glued
together directly, but rather joined via using smooth transition functions
of the form $(25)$ or $(26)$. In this particular case, the manifold
structure is $\mathcal{M}=\mathcal{M}^{-}\cup\mathcal{O}\cup\mathcal{M}^{+}$,
where $\mathcal{O}$ is a transition region with boundary hypersurfaces
$\Sigma_{\pm}=\partial\mathcal{O}\cup\partial\mathcal{\mathcal{M}^{\pm}}$,
and condition $(31)$ takes the form

\begin{equation}
e_{ab}^{\pm}=f^{\pm ab}=0,\:C_{\;ef}^{\pm a}=0.
\end{equation}
As may be noticed, this condition is fulfilled on $\Sigma_{\pm}$
due to the fact that the deformation fields $e_{ab}^{\pm}$ and $f^{\pm ab}$
have been chosen to be local tensor fields with the property that
all their components possess compact supports in $\mathcal{M}\text{\ensuremath{\backslash\mathcal{M}^{\pm}}}$.
To be more precise, based on the fact that e.g. the choice $e_{ab}^{\pm}=\chi^{\pm}\mathtt{e}_{ab}^{\pm}$
can always be made in the given context, where $\chi^{\pm}$ are smooth
transition functions of the form $(25)$ in which the constant $x_{0}$
is replaced by constants $x_{\pm}$ and $\mathtt{e}_{ab}^{\pm}$ are
continuous, at least twice differentiable tensor fields, it is clear
from the very outset that condition $(33)$ and therefore either condition
$(29)$ or $(30)$ are met as well on $\Sigma_{\pm}$. 

The same line of argument can be used to handle a variety of smooth
geometric transitions, i.e. to include partitions $\mathcal{M}=\mathcal{M}_{1}\cup\mathcal{O}_{1,2}\cup\mathcal{M}_{2}\cup...\mathcal{M}_{n-1}\cup\mathcal{O}_{n-1,n}\cup\mathcal{M}_{n}$
of the ambient manifold $\mathcal{M}$, where the $\mathcal{O}_{k,k+1}$
are transition regions connecting the four-dimensional Lorentzian
manifolds $\mathcal{M}_{k}$ and $\mathcal{M}_{k+1}$ with $k=1,2,...,n-1$.
This can be achieved by condsidering the deformation relations

\begin{equation}
g_{ab}=g_{ab}^{1}+e_{ab}^{1}=g_{ab}^{2}+e_{ab}^{2}=...=g_{ab}^{n}+e_{ab}^{n}
\end{equation}
and
\begin{equation}
g^{ab}=g_{1}^{ab}+f_{1}^{ab}=g_{2}^{ab}+f_{2}^{ab}=...=g_{n}^{ab}+f_{n}^{ab},
\end{equation}
which are given with respect to associated sequences of deformation
fields $e_{ab}^{1}$, $e_{ab}^{2}$, ...,$e_{ab}^{n}$ and $f_{1}^{ab}$,
$f_{2}^{ab}$,...,$f_{n}^{ab}$ that are chosen in such a way that
$e_{ab}^{(k)}=\chi_{k}\mathtt{e}_{ab}^{(k)}$ for $k=1,2,..,n$, where
each $\chi_{k}$ is a transition function of the form $(25)$ with
$x_{0}$ replaced by $x_{k}$ and $\mathtt{e}_{ab}^{(k)}$ is a continuous,
at least twice differentiable tensor field. The corresponding deformation
fields must be given such that they obey consistency relation $(18)$
and also 

\begin{equation}
e_{ab}^{(k)}=f_{(k)}^{ab}=0,\:C_{\;ef}^{(k)a}=0.
\end{equation}
By requiring this, however, it becomes clear that the field equations
of the theory will have a completely different form locally than globally.
Therefore, from a local point of view, the structure of the said equations
will change over time, just like that of the Einstein-Hilbert action.
This proves to be relevant for the action principle of the theory.

In the case that the ambient spacetime $(\mathcal{M},g)$ exhibits
a boundary $\partial M$ without edges or corners, this action is
given by

\begin{equation}
S[g]=\underset{M}{\int}R\omega_{g}+\int\limits _{\Sigma}^{\Sigma'}K\omega_{h},
\end{equation}
where $\Sigma$ and $\Sigma'$ are spacelike hypersurfaces and $\omega_{g}\equiv\sqrt{-g}d^{4}x$
is the four-volume element and $\omega_{h}\equiv\sqrt{h}d^{3}x$ is
the three-volume element of spacetime. If the same boundary $\partial M$
of the ambient spacetime, on the other hand, does indeed contain a
sharp edge or corner, its action alternatively can be specified by
Hayward's action \cite{brown1993quasilocal,hayward1993gravitational}

\begin{equation}
S[g]=\int\limits _{M}R\omega_{g}+\int\limits _{\Sigma}^{\Sigma'}K\omega_{h}+\int\limits _{\mathcal{B}}\tilde{K}\omega_{\gamma}+\int\limits _{\Omega}^{\Omega'}\sinh^{-1}\eta\omega_{q},
\end{equation}
where $\omega_{h}\equiv\sqrt{h}d^{3}x$, $\omega_{\gamma}\equiv\sqrt{-\gamma}dtd^{2}x$
and $\omega_{q}\equiv\sqrt{q}d^{2}x$ are volume forms associated
with the individual parts of the boundary $\partial M$ of $(\mathcal{M},g)$,
which consists of two spacelike hypersurfaces $\Sigma$ and $\Sigma'$
and a timelike hypersurface $\mathcal{B}$, which intersects the spacelike
hypersurfaces $\Sigma$ and $\Sigma'$ in $\Omega$ and $\Omega'$.
Here, the quantities $K$ and $\tilde{K}$ are extrinsic curvature
scalars and $\eta:=n_{a}u^{a}$ is a generally non-zero scalar parameter
originating from the fact that the boundary normals $n^{a}$ and $u^{a}$
are usually non-orthogonal in the given case.

Regardless of whether one or the other type of action is considered
in this context, it may happen that, in the course of a geometric
transition, the structure of the metric does not change momentarily
over time, whereas that of the Einstein equations does very well.
In such a case, the local metric and the ambient metric do coincide,
but neither the corresponding field equations nor the corresponding
action functionals $(37)$ or $(38)$ do so as well. 

Considering the simplest case of two local spacetimes $(\mathcal{M}^{\pm},g^{\pm})$,
the changes in the field equations can be described by equations $(23)$
and $(27)$. The reason why these changes have to be taken into account
here are the following: The tensor fields $\mathtt{e}_{ab}^{\pm}$
may be vanishing in $\mathcal{\mathcal{M}}\backslash\mathcal{\mathcal{M}^{\pm}}$,
so that it may happen that $e_{ab}^{\pm}\rightarrow0$ due to the
fact that $\mathtt{e}_{ab}^{\pm}\rightarrow0$ in $\mathcal{\mathcal{M}}\backslash\mathcal{\mathcal{M}^{\pm}}$.
Therefore, it must be expected that $C_{\;ef}^{\pm a}\rightarrow0$
applies in the event that $\chi^{\pm}\rightarrow0$, but not in the
event that $\mathtt{e}_{ab}^{\pm}\rightarrow0$, in which case $C_{\;ef}^{\pm a}\neq0$
rather applies in general. Therefore, it may occur that the local
metric and the ambient metric coincide, but not the corresponding
curvature fields. 

As a result, the structures of the action and the field equations
of the theory may change, but those of the local metrics may not.
In the case that an ambient spacetime $(\mathcal{M},g)$ with Lorentzian
manifold $\mathcal{M}=\mathcal{M}_{1}\cup\mathcal{O}_{1,2}\cup\mathcal{M}_{2}\cup...\mathcal{M}_{n-1}\cup\mathcal{O}_{n-1,n}\cup\mathcal{M}_{n}$
is given, which exhibits a boundary $\partial M$ without edges or
corners, the change of the field equations in the course of the $k$-th
geometric transition can straightforwardly be determined to be

\begin{equation}
G_{ab}^{(k)}+\rho_{ab}^{(k)}=8\pi T_{ab},
\end{equation}
where $\rho_{ab}^{(k)}=\psi_{ab}^{(k)}-\frac{1}{2}g_{ab}^{(k)}(f^{(k)cd}R_{cd}^{(k)}+f^{(k)cd}E_{cd}^{(k)})-\frac{1}{2}e_{ab}^{(k)}(R^{(k)}+f^{(k)cd}R_{cd}^{(k)}+g^{(k)cd}E_{cd}^{(k)}+f^{(k)cd}E_{cd}^{(k)})$
with $\psi_{ab}^{(k)}=E_{ab}^{(k)}-\frac{1}{2}g_{ab}^{(k)}(g^{(k)cd}E_{cd}^{(k)})$
applies in the given context. Consequently, using the definition $\tau_{ab}^{(k)}:=T_{ab}-T_{ab}^{(k)}$,
one obtains the deformed field equations
\begin{equation}
\rho_{ab}^{(k)}=8\pi\tau_{ab}^{(k)}.
\end{equation}
Also the change of the Einstein-Hilbert action can be determined step
by step in such a case. Assuming for this purpose that $\text{\text{\ensuremath{\mathbf{1}}}}+e^{i}$
is the matrix representation of the object $\delta_{b}^{\;a}+e_{b}^{(i)\:a}$
for the $i$-th spacetime partition $(\mathcal{M}_{i},g^{i})$, the
relation $\vert X\vert=e^{\ln\vert X\vert}=e^{tr\ln X}=1+\overset{\infty}{\underset{m=1}{\sum}}\frac{\left(tr\ln X\right)^{m}}{m!}$
between the determinant and the trace of a matrix $X$ can be set
up in order to obtain the identity $\sqrt{-g}=(1+\varphi^{i})\sqrt{-g^{i}}$,
where $\varphi^{i}=\overset{\infty}{\underset{m=1}{\sum}}\frac{\left(tr\ln(\text{\text{\ensuremath{\mathbf{1}}}}+e^{(i)})\right)^{m}}{2^{m}m!}$
applies by definition. Moreover, using the fact that one can always
decompose the lapse function $N$ of the ambient spacetime with respect
to the lapse function $N^{i}$ of the $i$-th local background such
that $N=N^{i}+e_{00}^{(i)}$, the result obtained implies that $\sqrt{h}=(1+\varphi^{i})\frac{N^{i}}{N^{i}+e_{00}^{(i)}}\sqrt{h^{i}}=(1+\varphi^{i})\frac{1}{1+\frac{e_{00}^{(i)}}{N^{i}}}\sqrt{h^{i}}=:(1+\varphi^{i})(1+\psi^{i})\sqrt{h^{i}}$,
where $\psi^{i}=\overset{\infty}{\underset{n=1}{\sum}}\left(-\frac{e_{00}^{(i)}}{N^{i}}\right)^{n}$holds
by definition. In addition, the extrinsic curvature tensor $K_{ab}$
of $(\mathcal{M},g)$ can be decomposed with respect to the $i$-th
extrinsic curvature tensor $K_{ab}^{i}$ of the $i$-th local spacetime
$(\mathcal{M}_{i},g^{i})$ in the form $K_{ab}=K_{ab}^{i}+\kappa_{bd}^{i}$.

Consequently, in the case of a boundary $\partial M$ without edges
or corners, the corresponding action decomposes according to the rule

\[
S[g]\equiv S[g^{1}]+\Sigma[g^{1},e^{1},f^{1}]=...=S[g^{n}]+\Sigma[g^{n},e^{n},f^{n}])=
\]

\begin{equation}
=\frac{1}{n}\left\{ S[g^{1}]+\Sigma[g^{1},e^{1},f^{1}]+...+S[g^{n}]+\Sigma[g^{n},e^{n},f^{n}])\right\} ,
\end{equation}
where
\begin{equation}
S[g^{i}]=\int\limits _{M_{i}}R_{i}\omega_{g^{i}}+\int\limits _{\Sigma_{i}}^{\Sigma_{i}'}K_{i}\omega_{h^{i}}
\end{equation}
and
\begin{align}
\Sigma[g^{i},e^{i},f^{i}] & =\int\limits _{M_{i}}\chi^{i}R_{i}\omega_{g^{i}}+\int\limits _{\Sigma_{i}}^{\Sigma_{i}'}(\varphi^{i}+\psi^{i}+\varphi^{i}\psi^{i})K_{i}\omega_{h^{i}}+\\
+ & \int\limits _{M_{i}}(1+\varphi^{i})(g_{i}^{bd}E_{bd}^{i}+f_{i}^{bd}R_{bd}^{i}+f_{i}^{bd}E_{bd}^{i})\omega_{g^{i}}+\nonumber \\
+ & \int\limits _{\Sigma_{i}}^{\Sigma_{i}'}(1+\varphi^{i})(1+\psi^{i})(g_{i}^{bd}\kappa_{bd}^{i}+f_{i}^{bd}K_{bd}^{i}+f_{i}^{bd}\kappa_{bd}^{i})\omega_{h^{i}}+\nonumber \\
+ & \int\limits _{M\backslash M_{i}}(1+\varphi^{i})(R_{i}+g_{i}^{bd}E_{bd}^{i}+f_{i}^{bd}R_{bd}^{i}+f_{i}^{bd}E_{bd}^{i})\omega_{g^{i}}+\nonumber \\
+ & \int\limits _{\Sigma\backslash\Sigma_{i}}^{\Sigma\backslash\Sigma_{i}'}(1+\varphi^{i})(1+\psi^{i})(K_{i}+g_{i}^{bd}\kappa_{bd}^{i}+f_{i}^{bd}K_{bd}^{i}+f_{i}^{bd}\kappa_{bd}^{i})\omega_{h^{i}}\nonumber 
\end{align}
applies for the $i$-th part of the action. Thus, it can be seen that
the action of the ambient spacetime $(\mathcal{M},g)$ can be decomposed
into a system of 'subactions' $S[g^{i}]+\Sigma[g^{i},e^{i},f^{i}]$,
whose variation with respect to $g_{i}^{ab}$ possibly leads to modifications
of the 'standard' field equations obtained from a variation of $S[g^{i}]$
with respect to $g_{i}^{ab}$. 

A similar, but slightly more complicated decomposition relation is
also obtained in the case of a boundary $\partial M$ with edges or
corners, which is consistent not least due to the fact that Hayward's
action has been shown to be additive in a generalized sense \cite{brill1994gravitational}.
The associated formalism therefore allows one to add up the Einstein-Hilbert
actions of spacetimes with non-smooth boundaries and different topologies
and causal structures. 

However, it is important to note in this context that all modifications
to the Einstein's field equations and the Einstein-Hilbert action
need to be consistent with the geometric structure of the ambient
spacetime $(\mathcal{M},g)$. This point marks an important difference
to alternative multi-metric theories of gravity treated in literature
for which a priori no such correspondence is required \cite{de2014massive,fierz1939relativistic,hassan2012bimetric,isham1971nonlinear}.

This proves to be a very important point in that deformations of spacetime
metrics do not always have to lead to physically meaningful results.
Consequently, it is important to ensure that the respective fields
are chosen in a meaningful way. In order to ensure this and to treat
models of physical interest, it generally proves to be useful to consider
only deformation fields, which allow the fulfillment of suitable energy
conditions \cite{hawking1973large} on $(\mathcal{M},g)$. By requiring
this, it is ensured that the resulting confined stress-energy tensor
is well-defined from a physical point of view. Moreover, it is ensured
that the same conditions locally hold on $(\mathcal{M}^{\pm},g^{\pm})$. 

Anyway, after this has now been clarified, it remains to be discussed
what advantages working with the deformation approach presented in
this section has over working with the thin shell formalism presented
in the previous section. 

One of the main advantages of working with the deformation approach
is more general and versatile than the thin shell formalism and and
other closely related approaches to the subject, such as, in particular,
Penrose's cut-and-paste method. This is not least because it allows
the metrics and curvature fields of pairs of local spacetimes to be
smoothly deformed into each other via introducing smooth transition
functions instead of step functions. In this context, as it turns
out, the main advantage of the geometric deformation approach compared
to thin-shell formalism is that it cannot fail in the sense that,
in principle, a smooth geometric transition always exists for arbitrary
spacetime pairs. In contrast, the gluing of arbitrary spacetime pairs
is not always possible. 

In addition, as shall be explained in greater detail in the next section
on the basis of concrete geometric examples, the deformation approach
allows for a more careful handling of the subject in the sense that
it allows the treatment of problems where the thin shell method is
expected to lead to distributionally ill-defined terms, which cannot
be properly treated from a mathematical point of view. It turns out,
however, that the thin-shell formalism can be extended using Colombeau's
theory of generalized functions in order to enable a mathematically
rigorous treatment of the problematic terms mentioned, whereas all
extensions of the formalism mentioned prove to be completely consistent
with the geometric deformation approach.

Furthermore, in contrast to the thin shell formalism, the geometric
deformation approach also allows new solutions of the field equations
to be constructed using transformations that leave the geometric character
of the background metric unchanged, but lead to a new ambient spacetime
or even classes of ambient spacetimes.

Last but not least the deformation formalism includes the perturbative
approach to general relativity as a special case. Not least for this
reason, it allows one to weaken conditions $(28-33)$, which are designed
to meet the previously discussed junction conditions in various cases,
in a perturbative sense, so that they are no longer exact, but only
approximately valid, i. e. up to higher orders in a fixed parameter
or systems of parameters. 

All this will be explained in the next section using concrete geometric
models. For the sake of simplicity, however, only simple models will
discussed, which can be obtained by specifying a suitable deformation
of a (usually highly symmetric) background geometry.

\section{Geometric Deformations, Thin Shells and distributional Metrics of
Spacetime }

In the previous section, it was argued that the geometric deformation
framework is more general than the thin shell formalism and, moreover,
can be used (in combination with Colombeau's theory of generalized
functions) to extend said formalism to such an extent that the gluing
of spacetime metrics with low regularity becomes possible. Specifically,
it was stressed that spacetime metrics with components containing
delta functions can be glued together using the geometric deformation
framework. This shall now be demonstrated by some concrete geometric
examples, where - on the basis of the metric deformation formalism
discussed in the previous section - it will be made clear that certain
classes of distributional spacetimes are better suited to be glued
together than others. In due course, it will be made clear that the
thin shell formalism not only yields the exact same results as the
deformation approach, but rather emerges as a special case from this
approach. Furthermore, it will be made clear that deformation formalism,
in contrast to thin shell formalism, allows the gluing of arbitrary
pairs of local spacetimes by using suitable transition functions.

To deal with the points mentioned step by step, different classes
of spacetimes with deformed metrics shall be considered. The very
first of these classes will be the so-called generalized Gordon class
\cite{baccetti2012gordon,gordon1923lichtfortpflanzung,visser2010acoustic};
a class of spacetimes $(\bar{\mathcal{M}},\bar{g})$ with metrics
of the type 
\begin{equation}
\bar{g}_{ab}=g_{ab}+fn_{a}n_{b}.
\end{equation}
This special class of spacetimes can be obtained directly by deforming
the metric $g_{ab}$ of a given a background spacetime $(\mathcal{M},g)$,
using only two different objects: Some function $f$, whose form is
either known in advance or must be determined by solving the field
equations of the theory, and a smooth non-vanishing co-vector field
$n_{a}=g_{ab}n^{b}$, which shall be assumed to be timelike in relation
to the background metric $\ensuremath{g_{ab}}$, i.e. $g_{ab}n^{b}n^{b}<0$.
For the sake of simplicity, said vector field shall even be assumed
to be normalized with respect to the background metric, so that $g_{ab}n^{a}n^{b}=-1$
applies in the present context.

A well known example of a spacetime metric, which lies in the resulting
generalized Gordon class of metrics, is the so-called acoustic metric,
which plays an important role in describing deflections of light or
sound in bodies with different optical densities or acoustic properties
in both Special and General Relativity. There are, however, also other
important representatives of this class, many of which lie in a closely
related class of metrics, the so-called generalized conformal Gordon
class, which is a subclass of the generalized Gordon class with metrics
of the form

\begin{equation}
\bar{g}_{ab}=\Omega^{2}(g_{ab}+fn_{a}n_{b}).
\end{equation}
 Important representatives of this subclass have been studied within
the theory of so-called acoustic black holes and in the context of
analogue gravity; theories that aim to explain, among other things,
the geometric structure of acoustic black holes as well as electromagnetic
phenomena in linear media and the behavior of condensed matter models
in General Relativity or even more general theories of gravity \cite{barcelo2001analogue,fagnocchi2010relativistic,hossenfelder2017analogue,unruh1981experimental,visser2010acoustic}. 

Given this special class of metrics and associated spacetimes, one
may now ask the question of how two spacetimes of this class can be
joined with each other. To address this question, one may consider
a thin shell splitting and therefore make the specific choice $f=\theta f_{+}+(1-\theta)f_{-}$
for the function $f$ in $(44)$, where for the time being it shall
be assumed that $f_{\pm}$ are $C^{2}$-functions. In addition, it
may be required that $[f]=0$ holds on a spacelike hypersurface $\Sigma$
in spacetime. The resulting form of metric can then straightforwardly
be brought into a form of type $(1)$ by adding and subtracting a
term term of the form $\theta g_{ab}$, provided that the definitions
$g_{ab}^{\pm}\equiv g_{ab}+f_{\pm}n_{a}n_{b}$ are used in the present
context. But this makes clear that by this special splitting of the
function $f$, a splitting of the metric in the sense of the thin
shell formalism results. In a similar way, however, a decomposition
of the metric in the sense of the geometric deformation approach,
i.e. a splitting of the form $(15)$, can be obtained. To obtain such
a decomposition of the metric, one may simply add and subtract the
term $f_{\pm}n_{a}n_{b}$ in $(44)$, which yields $\bar{g}_{ab}=g_{ab}^{\pm}+e_{ab}^{\pm}$,
where the definitions $e_{ab}^{+}\equiv(1-\theta)(f_{-}-f_{+})n_{a}n_{b}$
and $e_{ab}^{-}\equiv\theta(f_{+}-f_{-})n_{a}n_{b}$ are used. Consequently,
as can be seen, both decompositions are equal, so that it becomes
clear - due to the fact that it is known that the thin-shell formalism
for $C^{2}$-metrics yields mathematically meaningful results - that
also the deformation approach must yield mathematically meaningful
results in the given case. 

More precisely, in view of the fact that condition $(18)$ can readily
be satisfied in a distributional sense by making an ansatz of the
form $f^{+ab}\equiv(1-\theta)\frac{f_{+}-f_{-}}{1+f_{+}-f_{-}}n^{a}n^{b}$
and $f^{-ab}\equiv\theta\frac{f_{-}-f_{+}}{1+f_{-}-f_{+}}n^{a}n^{b}$,
it can be seen that conditions $(28)$ and $(29)$ on $\Sigma$ are
automatically fulfilled due to the that $[f]=0$ must hold on that
very hypersurface. To additionally meet the stricter conditions listed
in $(31)$, it must be additionally required that $[\partial_{a}f]=0$
applies in the given context. However, due to the fact that there
is a great number of suitable choices for the functions $f_{\pm}$,
namely all, in relation to which either $[f]=0$ or $[f]=0$ and $[\partial_{a}f]=0$
applies. Concretely, the functions $f_{\pm}$ may be selected as $C^{\infty}$-functions
with compact supports in $M_{\mp}$, so that it becomes possible to
construct local spacetimes in the sense of section two of this work.
As can be seen, the fulfillment of the junction conditions of the
theory does not cause any problems in this context.

A problem that arises on the other hand is that it can happen that
a stress-energy tensor of the form $(9)$ does not allow the fulfillment
of relevant energy conditions of the theory or in special cases (as
in the case of the dominant energy condition) does not even allow
a mathematically meaningful formulation of the mentioned conditions
at all. But apart from this particular drawback, the formalism produces
mathematically and physically meaningful results.

However, the situation changes drastically if the requirement that
$f_{\pm}(x)$ are (at least) $C^{2}$-functions is dropped, and it
becomes particularly problematic when the choice $f_{\pm}=f_{\pm}^{0}\delta$
is made, where $f_{\pm}^{0}(x)$ are are $C^{2}$-functions functions
and $\delta(x)$ is the Dirac delta distribution. In this particular
case, the association relation $\theta\delta\approx A\delta$, previously
considered in section one of this work, can be used to show that relation
$(44)$ takes the form $\bar{g}_{ab}=g_{ab}+f_{0}\delta n_{a}n_{b}$;
at least provided that the definition $f_{0}=Af_{+}^{0}+(1-A)f_{-}^{0}$
is used in the present context. The problem, which then arises in
this context, is the following: Given this special distributional
form of the metric, one is confronted, as already indicated in section
one, with the serious problem that the curvature of spacetime and
all associated quantities cannot simply be calculated by considering
products of the delta distribution. The reason for this is that such
an approach would lead to undefined 'squares' of the Dirac delta distribution,
which are ill-defined from a mathematical point of view. For this
reason, the thin shell formalism cannot lead to meaningful results
in this particular case.

However, there is a feasible way to deal with this issue, which is
to use Colombeau's theory of algebras of generalized functions. Given
a paracompact manifold $X$, the center of attention of this theory
is the so-called Colombeau algebra $\mathcal{G}(X)$ (or rather an
entire system of Colombeau algebras, as there are many of such algebras),
which is a commutative, associative and unital differential algebra
of manifold-valued generalized functions. As such, it is an algebra
consisting of one-parameter families of $C^{\infty}$-functions $(f_{\varepsilon}(x))_{\varepsilon\in(0,1]}$,
which have to meet certain growth conditions in the so-called regularization
parameter $\varepsilon$. To be more precise, $\mathcal{G}(X)$ results
from forming the quotient algebra $\mathcal{E}_{m}(X)/\mathcal{N}(X)$
of the algebra of nets of moderate functions $\mathcal{E}_{m}(X)=\{(f_{\varepsilon})_{\varepsilon}\in C^{\infty}(X)^{(0,1]}:\forall K\subset\subset X\:\forall P\in\mathcal{P}(M)\:\exists l\:\underset{x\in K}{\sup}\vert Pf_{\varepsilon}(x)\vert=O(\varepsilon^{-l})\}$
by the ideal of nets of so-called negligible functions $\mathcal{N}(X)=\{(f_{\varepsilon})_{\varepsilon}\in C^{\infty}(X)^{(0,1]}:\forall K\subset\subset X\:\forall m\:\forall P\in\mathcal{P}(M)\:\underset{x\in K}{\sup}\vert Pf_{\varepsilon}(x)\vert=O(\varepsilon^{m})\}$,
where, in this context, $\mathcal{P}(X)$ denotes the space of all
linear differential operators on the manifold $X$. It may be noted
that $\mathcal{G}(X)$ contains the vector space of Schwartz distributions
as a linear subspace, and the space of smooth functions as a faithful
subalgebra.

Working with Colombeau algebras of generalized functions has a decisive
advantage over working directly with distributions: It allows one
to perform nonlinear operations on generalized functions, which result
in well-defined expressions coinciding with distributions in the limit
$\varepsilon\rightarrow0$. Therefore, by considering Colombeau algebras,
it becomes possible to treat mathematical problems that cannot be
treated in the standard theory of Schwartz distributions. In particular,
it becomes possible to perform nonlinear operations on a so-called
strict delta net $(\delta_{\varepsilon})_{\varepsilon\in(0,1]}\in C^{\infty}(\bar{M})^{(0,1]}$,
which converges to the delta distribution in the limit $\varepsilon\rightarrow0$
and thus allows for a regularization of the delta distribution \cite{colombeau2000new,colombeau2011elementary,grosser2001geometric}.
This offers the possibility to work with a delta sequence $\delta_{\varepsilon}$
instead of directly with the delta distribution and thus calculate
undefined products of the delta distribution in a mathematically rigorous
way. However, the situation is subtle: Different regularizations of
the delta distribution can lead to different results, which may lead
to different physical interpretations of the subject. To ensure that
a physically meaningful result is obtained in the end, one must therefore
be careful when selecting a preferred regularization by hand, as can
be illustrated very well by the example of the distributional Gordon
metric discussed above. 

To illustrate this, it is sufficient to consider a simple geometric
example. As a basis for considering such a simple example, the simplifying
assumption shall be made that a covariantly constant timelike normal
vector field $n^{a}$ exists on the background spacetime $(\mathcal{M},g)$,
i.e. a vector field with the properties $\nabla_{a}n^{b}=0$ and $g_{ab}n^{a}n^{b}=-1$.
Using this vector field to set up (in somewhat sloppy notation) a
relation of the form 
\begin{equation}
\bar{g}_{ab}^{\varepsilon}=g_{ab}+f_{0}\delta_{\varepsilon}\cdot n_{a}n_{b},
\end{equation}
where $f_{0}$ is a continuous, at least twice differentiable function
and $(\delta_{\varepsilon})_{\varepsilon}$ is a delta sequence that
coincides with the delta distribution in the limit $\varepsilon\rightarrow0$.
To treat a particularly easy example, it shall be assumed that the
delta sequence $\delta_{\varepsilon}$ and the function $f_{0}$ are
specified in such a way that $\nabla_{a}\delta_{\varepsilon}=-n_{a}\dot{\delta_{\varepsilon}}$
and $\nabla_{a}f_{0}=-n_{a}\dot{f_{0}}$ applies. From $(46)$ then
follows directly

\begin{equation}
\bar{g}_{ab}=\underset{\varepsilon\rightarrow0}{\lim}\:\bar{g}_{ab}^{\varepsilon}=g_{ab}+f_{0}\delta\cdot n_{a}n_{b}.
\end{equation}
Given this form of the metric, the inverse metric can be obtained
by making an ansatz of the form 
\begin{equation}
\bar{g}_{\varepsilon}^{ab}=g^{ab}+h_{0}\delta_{\varepsilon}\cdot n^{a}n^{b},
\end{equation}
where $h_{0}$ is some function that has to be determined with respect
to $f_{0}$ via solving 

\begin{equation}
\underset{\varepsilon\rightarrow0}{\lim}\:\bar{g}_{ab}^{\varepsilon}\bar{g}_{\varepsilon}^{bc}=\delta_{a}^{\;c}.
\end{equation}
Relation $(48)$ then yields an inverse metric of the type 
\begin{equation}
\bar{g}^{ab}=\underset{\varepsilon\rightarrow0}{\lim}\:\bar{g}_{\varepsilon}^{ab}=g^{ab}+h_{0}\delta\cdot n_{a}n_{b},
\end{equation}
 whereas it turns out that the value of $h_{0}$ will depend on the
choice of regularization of the delta distribution. For the possibility
of selecting a particular regularization by hand leads in this context
to ambiguities or, to put it more precisely, to different solutions
for relations $(49)$ and $(50)$. For example, there are regularizations
of the delta distribution, which lead to the result $\delta^{2}\approx0$,
while there are also other types of regularizations, which instead
lead to the result $\delta^{2}\approx c\delta$ \cite{grosser2001geometric,koh1992distributions,li2007review,li2014defining,michael1992multiplication}.
Both of these types of regularizations can be used to handle the nonlinear
operations on distributions \footnote{Note that there are many ways of modeling $\delta^{2}$ in Colombeau's
theory, but these will not be of further relevance at this point,
because they are not suitable for solving the problem at hand.}, which are necessary to solve relation $(49)$, where both relations
are equally correct from a purely mathematical point of view. This
is why, from a physics standpoint, the question arises as to which
of the two choices of regularization is the most reasonable and whether
there are other useful regularizations that allow to solve the problem
at hand or not. Ultimately, from the point of view of Colombeau's
theory of generalized functions, there is no silver bullet to solve
relations like $(49)$ and to derive from them, by means of solving
$(50)$, the form of the inverse metric $\bar{g}^{ab}$; in order
get then in the position to determine the form of the Levi-Civita
connection $\bar{\Gamma}_{\;bc}^{a}$ and the field equations of the
theory, which, in the given context, take the form 
\begin{equation}
\bar{G}_{ab}\approx8\pi\bar{T}_{ab},
\end{equation}
or equivalently 
\begin{equation}
\bar{R}_{ab}\approx8\pi\left(\bar{T}_{ab}-\frac{1}{2}\bar{g}_{ab}\bar{T}\right),
\end{equation}
where the LHS of $(52)$ is given by $\bar{R}_{ab}=R_{ab}+E_{ab}$
with a deformed Ricci tensor distribution of the form $E_{ab}=2\nabla_{[c}C_{\,b]a}^{c}+2C_{d[c}^{c}C_{b]a}^{d}$. 

However, as may be realized, making an optimal choice of delta regularization
is also important from a mathematical point of view, since such a
choice plays an essential role in the solution of the generalized
field equations of the theory for the given type of metric deformation.
For in order to find a solution of the mentioned equations, different
problematic distributional products must be determined, i.e. different
powers of the delta distribution as well as different powers of its
derivative. However, the calculation of such products is, if at all,
only possible for a suitable choice of regularization of the delta
distribution. Regardless of the concrete choice to be made, however,
one finds in the given case that

\begin{equation}
C_{d[c}^{c}C_{b]a}^{d}\approx0
\end{equation}
is valid, which implies that 
\begin{equation}
E_{ab}\approx2\nabla_{[c}C_{\,b]a}^{c}.
\end{equation}
Unfortunately, despite the validity of this identity, it generally
proves to be difficult to solve the remaining part of the field equations
even when using Colombeau's framework of generalized functions. The
problem here is not only that it is difficult to find solutions of
$(51)$ and $(52)$, respectively, but that it must be expected that
there are several solutions for the problem under consideration, whereby
it is unclear which one is the most interesting and most suitable
from a physical point of view. As shown in the rear part of this section,
there are, however, also cases in which the above mentioned regularizations
allow one to join pairs of distributional generalized Gordon spacetimes
in a mathematically rigorous way.

Notwithstanding that, from a pragmatic point of view, it generally
proves to be advantageous to continue to consider generalized Gordon
metrics, which are $C^{2}$, when gluing them together. For even with
the simplest types of distributional metrics of this form, it may
occur that the thin shell formalism reaches its limits - even when
using Colombeau's theory of generalized functions. On the other hand,
this cannot happen in the $C^{2}$-case. Here Colombeau's theory is
only needed to set up problematic energy conditions (such as the dominant
energy condition) for the stress-energy tensors of the form $(9)$,
which are undefined in the Schwartz theory of distributions. 

But when it comes to demonstrating that spacetimes with metrics containing
a delta term can be unambiguously glued together, it is advisable
to consider another class of deformed spacetimes whose representatives
lead to simpler curvature expressions and therefore to a simpler structure
of the field equations.

One class of local geometric deformations, for which this is actually
the case, is the class of so-called generalized Kerr-Schild deformations;
a class, whose representatives are metrics of the type

\begin{equation}
\bar{g}_{ab}=g_{ab}+fl_{a}l_{b},
\end{equation}
 which are given with respect to a lightlike geodesic co-vector field
$l_{a}=g_{ab}l^{b}$ that meets the conditions $g_{ab}l^{a}l^{b}=0,\ \left(l\nabla\right)l^{a}=0$
and $\bar{g}_{ab}l^{a}l^{b}=0,\ (l\bar{\nabla})l^{a}=0$. 

The generalized Kerr-Schild class is a family of solutions that encompass
a considerably large class of geometric models that are of interest
to General Relativity, such as all stationary geometries that are
in the Kerr-Newman family of spacetimes and, in addition, all dynamical
radiation fluid spacetimes lying in the even more general Bonnor-Vaidya
family. Moreover, it includes various models with cosmological horizons,
like for instance Kottler alias Schwarzschild-de Sitter spacetime
and its generalizations.

As in the case of the generalized Gordon class of metrics, the problem
of gluing pairs of generalized Kerr-Schild spacetimes can be treated
in the given case by considering a thin shell splitting of the form
$f=\theta f_{+}+(1-\theta)f_{-}$ in $(55)$ and requiring that $[f]=0$
on a lightlike hypersurface $\Sigma$. In this case, the metric $\bar{g}_{ab}$
of the ambient spacetime can straightforwardly be decomposed in such
a way that it takes the form $(1)$. Alternatively, one may add and
subtract a term of the form $f_{\pm}l_{a}l_{b}$ in $(55)$ to obtain
a splitting of the form $(15)$, which yields $\bar{g}_{ab}=g_{ab}^{\pm}+e_{ab}^{\pm}$,
where the definitions $e_{ab}^{+}\equiv(1-\theta)(f_{-}-f_{+})l_{a}l_{b}$
and $e_{ab}^{-}\equiv\theta(f_{+}-f_{-})l_{a}l_{b}$ are used. Making
then an ansatz of the form $f_{\pm}=f_{\pm}^{0}\delta$, where $f_{\pm}^{0}(x)$
are are $C^{2}$-functions functions and $\delta(x)$ is the Dirac
delta distribution, one obtains the result 
\begin{equation}
\bar{g}_{ab}=g_{ab}+f_{0}\cdot\delta l_{a}l_{b},
\end{equation}
provided that the definition $f_{0}=Af_{+}^{0}+(1-A)f_{-}^{0}$ is
used in the present context. In order to be able to perform those
nonlinear operations on $\bar{g}_{ab}$ needed in order to set up
the inverse metric, the Levi-Civita connection $\bar{\Gamma}_{\;bc}^{a}$
and Einstein's field equations, which have again the form $(51)$
and $(52)$, respectively, also in the given case a strict delta net
$(\delta_{\varepsilon})_{\varepsilon\in(0,1]}\in C^{\infty}(\bar{M})^{(0,1]}$
shall be condidered, which converges to the delta distribution in
the limit $\varepsilon\rightarrow0$. The associated regularized generalized
Kerr-Schild metric 
\begin{equation}
\bar{g}_{ab}^{\varepsilon}=g_{ab}+f_{0}\delta_{\varepsilon}\cdot l_{a}l_{b}
\end{equation}
can then be used to obtain the form of the inverse metric via solving
relation $(49)$ with respect to the ansatz 

\begin{equation}
\bar{g}_{\varepsilon}^{ab}=g^{ab}-f_{0}\delta_{\varepsilon}\cdot l^{a}l^{b}
\end{equation}
regularized generalized inverse Kerr-Schild metric. This yields the
result

\begin{equation}
\bar{g}^{ab}=\underset{\varepsilon\rightarrow0}{\lim}\:\bar{g}_{\varepsilon}^{ab}=g^{ab}-f_{0}\delta\cdot n_{a}n_{b},
\end{equation}
where, in contrast to the previous case of the Gordon class of metrics,
it turns out that the form of $(59)$ is independent of the choice
of regularization of the delta distribution.

As shown in \cite{taub1981generalised}, the geometric structure of
the deformed field equations is particularly simple in the given case
of the generalized Kerr-Schild class. More specifcally, the mixed
Einstein tensor $\bar{G}_{\;b}^{a}$ of the ambient metric $\bar{g}_{ab}$
turns out to be linear in the profile function $f$. Moreover, as
shown in \cite{huber2020distributional}, also the Einstein tensor
with lowered and raised indices are linear in $f\equiv f_{0}\cdot\delta$
if the geometric constraints

\begin{equation}
\bar{\nabla}_{[a}l_{b]}=\nabla_{[a}l_{b]}=0,\;(l\bar{\nabla})f=(l\nabla)f=0
\end{equation}
are met. To see this, one may use the fact that there holds 
\begin{equation}
C_{\:bc}^{a}\approx\frac{1}{2}\nabla_{b}(fl^{a}l_{c})+\frac{1}{2}\nabla_{c}(fl^{a}l_{b})-\frac{1}{2}\nabla^{a}(fl_{b}l_{c}),\;C_{\:ab}^{b}\approx0
\end{equation}
for the corresponding deformation tensor of the generalized Kerr-Schild
class, which relates the pair of covariant derivative operators $\bar{\nabla}_{a}$
associated with $\bar{g}_{ab}$ and $\nabla_{a}$ and associated with
$g_{ab}$. Using this result, one finds that the deformed Ricci tensor
with lowered indices reads 
\begin{equation}
\bar{R}_{ab}=R_{ab}+E_{ab},
\end{equation}
where $E_{ab}=\nabla_{c}C_{\:ab}^{c}+C_{\:ad}^{c}C_{\:cb}^{d}$ applies.
As it then turns out in this context, the conditions 
\begin{equation}
C_{\:ad}^{c}C_{\:cb}^{d}\approx0
\end{equation}
and 
\begin{equation}
\nabla_{c}C_{\:ab}^{c}l^{a}\approx0,\nabla_{c}C_{\:ab}^{c}l^{b}\approx0,\;E_{\;b}^{a}l_{a}l^{b}\approx0
\end{equation}
are met regardless of the choice of regularization of the delta distribution,
where the conditions listed in $(64)$ result from the consistency
condition $\bar{R}_{ab}=\bar{g}_{ac}\bar{R}_{\;b}^{c}$. Thus, using
Colombeau's theory of generalized functions in combination with the
geometric framework of local metric deformations, one finds that the
field equations of the theory are linear in $f=f_{0}\cdot\delta$.

From these results, it can be concluded that the generalized Kerr-Schild
framework is well suited to address the problem of gluing spacetimes
of low regularity. In particular, it is found that it is much more
straightforward to glue together distributional metrics belonging
to the generalized Kerr-Schild class than distributional metrics belonging
to the generalized Gordon class. From this, however, it can be concluded
that the choice of the type of local geometric deformation determines
how well the problem of gluing spacetimes with low regularity can
be treated in practice, which in turn is the reason why the methods
discussed in section two are useful for gluing spacetimes with low
regularity.

In order to highlight this point, a few concrete geometric examples
shall now be considered, whereas the main focus shall be placed directly
on representatives of the generalized Kerr-Schild class. This makes
sense not least because this class provides the best known examples
of pairs of distributional spacetimes that can be glued together.
The best known examples are gravitational shock wave geometries, whose
importance for the thin shell formalism was recognized long ago \cite{taub1980space}
(although the treatment of mathematically problematic expressions
has not received the necessary attention at the time). The focus in
this context shall be on gravitational shock wave geometries in black
holes and cosmological backgrounds. These geometries, which all were
found by using Penrose's cut-and-paste alias scissors-and-paste procedure,
that is, a method for gluing spacetimes along lightlike hyperfaces,
characterize the fields of spherical shock waves caused by a massless
particle moving at the speed of light along the corresponding event
or cosmological horizons. The most famous representatives of this
class are the geometries of Dray and \cite{dray1985gravitational},
Sfetsos \cite{sfetsos1995gravitational} and Lousto and Sanchez \cite{lousto1989ultrarelativistic},
which characterize the gravitational fields of spherical shock waves
in Schwarzschild, Reissner-Nordström and Kottler alias Schwarzschild-de
Sitter backgrounds. 

All of these geometries have in common that their line elements can
be written down in the form

\begin{equation}
ds^{2}=2B^{2}f_{0}\delta dU^{2}-2B^{2}dUdV+r^{2}(d\theta^{2}+\sin^{2}\theta d\phi^{2})
\end{equation}
where $\delta=\delta(U)$ is Dirac's delta distribution and $B=B(UV)$
is a function whose explicit form depends of whether background spacetime
is Schwarzschild, Reissner-Nordström or Kottler. Thus, it can be concluded
that the metrics 
\begin{equation}
\bar{g}_{ab}=g_{ab}+2B^{2}f_{0}\delta l_{a}l_{b},
\end{equation}
corresponding to these line elements belong to the generalized Kerr-Schild
classes of the respective backgrounds, where in each case one has
$l_{a}=g_{ab}l^{b}=-dU_{a}$. Accordingly, given the fact that one
can always choose $f_{0}=f_{0}(\theta,\phi)$ with $f_{0}=Af_{+}^{0}+(1-A)f_{-}^{0}$
in this context, it becomes clear that the junction conditions $(32)$
are met if it is required that $\partial_{V}B\vert_{U=0}=\partial_{V}r\vert_{U=0}=0$.
As a result, the Einstein tensor of the corresponding classes of geometries
takes the form $G_{\;b}^{a}=(\Delta-c)f_{0}\delta l^{a}l_{b}$ and
thus characterizes the geometric field of a null fluid source. 

The validity of this result cannot be deduced from thin shell formalism
alone; it requires geometric deformation theory to make it possible.
This can be concluded from the fact that in the past, on the basis
of careless application of Penrose's method, which according to \cite{dray1985gravitational}
gives the same results as the thin shell formalism (except for a single
not particularly relevant term), the authors of the above-mentioned
works came to the erroneous conclusion that despite the validity of
$\partial_{V}B\vert_{U=0}=\partial_{V}r\vert_{U=0}=0$, the field
equations should contain ill-defined 'delta-square' terms. As it turns
out, however, the deformed field equations of the generalized Kerr-Schild
class do not contain such terms after all, but lead to a single differential
equation for the reduced profile function of the form

\begin{equation}
(\Delta-c)f_{0}=2\pi b\delta,
\end{equation}
where $\delta\equiv\delta(\cos\theta-1)$ is Dirac's delta distribution
and $b$ and $c$ are constants, whereas $c$ is given by $c=2r_{+}(\kappa-\Lambda r_{+})$
in the Schwarzschild-de Sitter case, $c=2r_{+}\kappa$ in the Reissner-Nordström
case and by $c=1$ in the Schwarzschild case. 

The resulting equation can be solved by expanding the reduced profile
function on the left hand side and the delta function on the right
hand side simultaneously in Legendre polynomials. Using here the fact
that $\delta(x)=\overset{\infty}{\underset{l=0}{\sum}}(l+\frac{1}{2})P_{l}(x)$,
one obtains the solution 
\begin{equation}
f_{0}(\theta)=-b\overset{\infty}{\underset{l=0}{\sum}}\frac{l+\frac{1}{2}}{l(l+1)+c}P_{l}(\cos\theta)
\end{equation}
by solving the corresponding eigenvalue problem. As was shown in \cite{sfetsos1995gravitational},
however, it is quite possible to find another representation for the
function $f$, which is fully consistent with the thin shell formalism
discussed in section one of this work. 

Other examples of Kerr-Schild deformed local spacetimes with deformation
fields that have compact support in a single null hypersurface of
the geometry are pp-wave spacetimes. The perhaps most well-known models
in this regard are the spacetimes of Aichelburg and Sexl \cite{aichelburg1971gravitational}
and Lousto and Sanchez \cite{lousto1989ultrarelativistic,lousto1990curved},
which have in common that they are specified by a Brinkmann form that
contains a delta distribution and therefore has support only on a
single lightlike hypersurface of spacetime. For that reason, they
determine a local background geometry that coincides everywhere with
that of a spherically symmetric black hole spacetime except for one
single null hypersurface. 

Consequently, as it turns out, the deformation approach is not only
fully compatible with the thin shell formalism, but also shows the
treatability of problems that cannot be treated by the naive application
of the standard methods of said formalism. But this does not only
concern the gluing of distrubtional metrics: In contrast to the thin
shell formalism, the geometric deformation approach allows the smooth
gluing of arbitrary pairs of local spacetimes by using a suitable
set of transition functions.

This is because the thin shell formalism requires that only pairs
of spacetimes can be glued together for which the corresponding junction
conditions are fulfilled. On the other hand, the use of suitable transition
functions within the geometric deformation approach ensures that pairs
of local spacetimes can be glued together smoothly, i.e. without the
possibility of the existence of a singular confined stress-energy
tensor with a delta shock at the joint boundary hyperface of the spacetimes.

This can be easily made clear by considering a splitting of the form
$f=\chi_{+}f_{+}+\chi_{-}f_{-}$ either in $(44)$ or in $(55)$,
where the $\text{\ensuremath{\chi_{\pm}}}$ are functions of the form
$(25)$ which are zero in $M_{\pm}$ and $f_{\pm}$ are essentially
arbitrary functions, which, however, shall be assumed to be at least
$C^{2}$. Such a choice makes it possible to model local spacetimes
whose geometry changes continuously (and not instantly from one moment
to the next, as in the thin shell formalism) in such a way that a
given initial geometry transitions smoothly into a certain final geometry
of spacetime. Thus, in other words, choosing the function $f$ in
this way allows a smooth geometric transition between pairs of local
spacetimes $(\mathcal{M}^{\pm},g^{\pm})$.

To demonstrate this, the special case $f=\chi f_{0}$ shall be considered
for a generalized Kerr-Schild metric in $(55).$ More specifically,
the Bonnor-Vaidya family of spacetimes \cite{bonnor1970spherically}
shall be considered, whose metric, in the general rotating case, can
be read off the line element

\begin{align}
d\bar{s}^{2} & =-dv^{2}+2(dv-a\sin^{2}\theta d\phi)dr+\Sigma d\theta^{2}+\\
 & +\frac{(r^{2}+a^{2})\sin^{2}\theta}{\Sigma}d\phi^{2}+\frac{2Mr-e^{2}}{\Sigma}(dv-a\sin^{2}\theta d\phi)^{2},\nonumber 
\end{align}
where $\Sigma=r^{2}+a^{2}\cos^{2}\theta$ and $M=M(v)$, $e=e(v)$.
The energy-momentum tensor of this geometry consists of a null fluid
part and an additional part, i.e. $\bar{T}_{ab}=\varepsilon l_{a}l_{b}+2\vartheta\left(l_{(a}k_{b)}+m_{(a}\bar{m}_{b)}\right)+2\varsigma l_{(a}\bar{m}_{b)}+2\bar{\varsigma}l_{(a}m_{b)},$where
$\varepsilon=-\frac{2r\left(r\dot{M}-e\dot{e}\right)+a^{2}\sin\theta\left(r\ddot{M}-\dot{e}\dot{e}-e\ddot{e}\right)}{8\pi\Sigma^{2}}$,
$\vartheta=\frac{e^{2}}{8\pi\Sigma^{2}}$ and $\varsigma=\frac{-ia\sin\theta}{\sqrt{2}8\pi\Sigma^{2}}\left\{ \Sigma\dot{M}-2e\dot{e}\right\} $
with $\dot{M}:=\frac{dM}{dv}$ and $\dot{e}:=\frac{de}{dv}$. 

By considering a splitting of the mass and the charge functions into
constant and non-constant parts, i.e.$M(v)=M_{0}+m(v)$ and $e(v)=e_{0}+\mathfrak{e}(v)$
with $M_{0}=const.$ and $e_{0}=const.$, the metric associated with
line elment $(69)$ can be written in the form 
\begin{equation}
\bar{g}_{ab}=g_{ab}+fl_{a}l_{b},
\end{equation}
where $f=\frac{m+2e_{0}\mathfrak{e}+\mathfrak{e}^{2}}{\Sigma}$ provided
that $l_{a}=-dv_{a}+a\sin^{2}\theta d\phi_{a}$. Due to the fact that
the mixed Einstein tensor is linear in the profile function, the energy-momentum
tensor decomposes according to the rule $\bar{T}_{\;b}^{a}=T_{\;b}^{a}+\tau_{\;b}^{a}$,
where $T_{\;b}^{a}$ is the energy-momentum tensor of the Kerr-Newman
black hole background spacetime. The resulting deformed geometry of
spacetime may therefore be phyiscally interpreted as the gravitational
field of a Kerr-Newman black hole that accretes null radiation. 

Although $m(v)$ and $\mathit{\mathfrak{e}}(v)$ may in principle
be chosen arbitrarily, one may choose them to be of the form $m(v)=\psi(\frac{v}{v_{0}})m_{0}$
and $\mathit{\mathfrak{e}}(v)=\psi(\frac{v}{v_{0}})\mathit{e}_{0}$,
where $m_{0}=const.$ and $\mathit{e}_{0}=const$. The transition
function $\psi(\frac{v}{v_{0}})$ takes a value of zero for $v<0$,
a value of one for $v\geq1$ and is strictly increasing in the interval
$[0,1]$, so that the set of conditions given in $(33)$ is met and
it can therefore be concluded that the metric of spacetime coincides
locally with that of Kerr-Newman spacetime; a spacetime that describes
the electrovac gravitational field of a stationary axially symmetric
charged rotating black hole, which has, by necessity, a completely
different physical interpretation from the metric of a rotating Bonnor-Vaidya
spacetime. More specifically, the spacetime geometry at hand describes
how an initially given Kerr-Newman black hole geometry with 'degrees
of freedom' $(M_{0},e_{0},a)$ transitions smoothly into one with
different parameter values $(M_{0}+m_{0},e_{0}+\mathit{e}_{0},a)$,
so that it can be concluded that Bonnor-Vaidya spacetime characterizes
the gravitational field of an charged rotating black hole that accretes
null radiation over a finite period of time. 

As a consequence, it is found that the Bonnor-Vaidya model can always
be set up to predict the collapse of a null radiation field and its
absorption by a charged rotating black hole, which could even result
in the complete discharge of the black hole, whereas it is woth mentioning
that these results are in complete agreement with the famous black
hole uniqueness theorems \cite{carter1971axisymmetric,robinson1974classification,robinson1975uniqueness}.

Of course, one could also try to make another choice for the function
$f$ in $(70)$, which is in better agreement with the thin shell
formalism. In particular, one could choose $m(v)=\theta(v-v_{0})m_{0}$
and $\mathit{e}(v)=\theta(v-v_{0})\mathit{e}_{0}$, where $\theta(v-v_{0})$
represents the Heaviside step function $\theta(v-v_{0}):=\begin{cases}
\overset{0}{\underset{1}{\frac{1}{2}}} & \overset{v-v_{0}<0}{\underset{v-v_{0}>0}{v-v_{0}=0}}\end{cases}$. However, from a purely physical point of view, this would actually
be a very poor choice, since the resulting geometry would describe
the very unphysical case of a black hole that accretes material of
mass $m_{0}$ and charge $\mathit{e}_{0}$ within an infinitesimally
small instant of time, which is why it is more sensible to stick to
the above smooth description of the problem. Nevertheless, the given
choice also provides a well-defined example of a local geometry in
the previously introduced sense and the resulting geometric model
reveals the structure of the gravitational field of a black hole that
absorbs null radiation.

Now that this has been clarified, the next thing to be noted is that
the junction conditions of the thin shell formalism result as a special
case of the discussed smooth geometric framework if the limit is considered
where the size of the smooth transition region goes to zero. To see
this, one may consider a transition region $\mathcal{O}$ with length
(or time) scale $L$ and the generalized function 
\begin{equation}
\chi_{L}(x-x_{0})=\frac{1}{2}(1+\tanh\frac{x-x_{0}}{L})=\frac{1}{1+e^{\frac{x-x_{0}}{L}}}
\end{equation}
 which provides a smooth, analytic approximation of the step function,
converging exactly to said function in the limit $L\rightarrow0$,
i.e. $\theta(x-x_{0})=\underset{L\rightarrow0}{\lim}\chi_{L}(x-x_{0})$.
Using this definition as a basis for setting up the decomposition
relation 
\begin{equation}
g_{ab}=\underset{L\rightarrow0}{\lim}\chi_{L}g_{ab}^{+}+(1-\chi_{L})g_{ab}^{-}
\end{equation}
and, moreover, the fact that 
\begin{equation}
\frac{d\theta(x-x_{0})}{dx}=\underset{L\rightarrow0}{\lim}\frac{1}{L}\frac{1}{2+e^{\frac{x-x_{0}}{L}}+e^{\frac{x_{0}-x}{L}}}
\end{equation}
turns out be singular for $x=x_{0}$ and zero otherwise, one finds
that condition $(2)$ must be met in order to join pairs of local
spacetimes $(\mathcal{M}^{\pm},g^{\pm})$. In addition, using the
fact that the generalized function $\chi_{L}(x)$ provides a smooth
approximation of the Heaviside step functionand its derivative with
respect to $x$ a smooth approximation of the delta distribution,
steps analogous to those described in section one lead first to relations
$(5)$ and $(6)$ and then to relations $(7-9)$. By distinguishing
the non-lightlike and lightlike cases one then finds that junction
conditions $(10)$ and $(11)$ must hold in the first case, whereas
junction conditions $(12)$ or $(13)$ must hold in the latter case.
Consequently, it follows that the junction conditions of the theory
can be derived via using the smooth metric deformation framework presented
in section two, but under the premise that the size of the transition
region approaches zero in a suitable limit. However, since there is
a wide range of generalized functions that are associated with the
step function and whose first derivatives are associated with the
delta distribution in a distributional sense, it becomes clear that
the local geometric deformation approach discussed in section two
of this work generalizes the thin-shell formalism discussed in section
one.

The next point to note is that by considering suitable transition
functions - as already mentioned earlier at the beginning of section
three - the unambiguous gluing of generalized Gordon class spacetimes
becomes possible. To see this, one may consider a generalized Gordon
class metric of the type

\begin{equation}
\bar{g}_{ab}=g_{ab}+\underset{\varepsilon\rightarrow0}{\lim}f_{0}\delta_{\varepsilon}\cdot n_{a}n_{b},
\end{equation}
which is defined with respect to a timelike vector field $n^{a}=\frac{1}{\sqrt{2}}\left(l^{a}+\chi k^{a}\right)$,
where $\chi$ is a smooth transition function of the form $(25)$
whose support does nor intersect that of the delta distribution resulting
from the limit $\underset{\varepsilon\rightarrow0}{\lim}\delta_{\varepsilon}$.
Generalized Gordon spacetimes of this type can always be glued together,
since the field equations coincide with those of the generalized Kerr-Schild
class in those local spacetime domains in which the delta distribution
becomes singular. The reason for this is that the timelike vector
field $n^{a}$ becomes locally lightlike by the given choice, so that
the gravitational shock wave geometries considered above can be generalized
in a mathematically rigorous way. However, this makes it clear that
in principle it is possible to glue generalized Gordon spacetimes
together without having to calculate probematic powers of the delta
distribution and its derivatives, which is a challenge despite the
use of Colombeau's theory of generalized functions. 

In any case, the results obtained so far can certainly be generalized
in many ways. This can be demonstrated, for example, by generalizing
the metric associated with the line element $(70)$. This can be accomplished
by performing a null rescaling of the form $l_{a}\rightarrow\lambda l_{a}$,
$k_{a}\rightarrow\lambda^{-1}k_{a}$, which leaves the geometric structure
of the background metric $g_{ab}$ invariant, but changes the geometric
structure of the metric $\bar{g}_{ab}$ of the ambient spacetime $\bar{(\mathcal{M}},\bar{g})$.
This yields a more general class of solutions to Einstein's field
equations with a metric of the form 

\begin{equation}
\bar{g}_{ab}=g_{ab}+\lambda^{2}fl_{a}l_{b},
\end{equation}
 where the only condition that one may wish to impose in this context
is that $(l\nabla)\lambda=0$ and therefore $\lambda=\lambda(v,\theta,\phi)$
holds by definition, so that the resulting class of geometries still
belongs to the generalized Kerr-Schild class of Kerr-Newman spacetime
and the corresponding mixed field equations remain linear in $f$.
But, of course, that restriction is not a must by any means.

Another possibility to construct a new class of models from the one
given above is to use the fact that the null geodesic vector field
can be extended to an associated null geodesic frame $(l^{a},k^{a},m^{a},\bar{m}^{a})$
and then to perform a null rotation of the form $k_{a}\rightarrow k_{a},$
$m_{a}\rightarrow m{}_{a}+\xi k_{a},$ $l_{a}\rightarrow l_{a}+\xi\bar{m}_{a}+\bar{\xi}m_{a}+\vert\xi\vert^{2}k_{a}$,
which again leaves the geometric structure of the background metric
$g_{ab}$ invariant, but changes the geometric structure of the metric
$\bar{g}_{ab}$ of the ambient spacetime $\bar{(\mathcal{M}},\bar{g})$.
In this way, once again a more general class of solutions to Einstein's
field equations is obtained, whose metric is of the form 
\begin{align}
\bar{g}_{ab} & =g_{ab}+fl_{a}l_{b}+2f\xi l_{(a}\bar{m}_{b)}+2f\bar{\xi}l_{(a}m_{b)}+\\
 & +2f\vert\xi\vert^{2}\left(l_{(a}k_{b)}+m_{(a}\bar{m}_{b)}\right)+f\xi^{2}\bar{m}_{a}\bar{m}_{b}+f\bar{\xi}^{2}m_{a}m_{b}+\nonumber \\
 & +2f\xi\vert\xi\vert^{2}k_{(a}\bar{m}_{b)}+2f\bar{\xi}\vert\xi\vert^{2}k_{(a}m_{b)}+f\vert\xi\vert^{4}k_{a}k_{b}.\nonumber 
\end{align}
Consequently, models of any complexity can be constructed by repeatedly
applying a combination of null rescalings and null rotations. Therefore,
it is actually quite straightforward to construct another, more general
type of ambient spacetime $\bar{(\mathcal{M}},\bar{g})$ with metric

\begin{equation}
\bar{g}_{ab}=g_{ab}+e_{ab}
\end{equation}
from an ambient spacetime with metric form $(70)$, whose geometry
coincides locally with that of Kerr-Newman spacetime $(\mathcal{M},g)$.
It may be noted that the concrete choice of the local background metric
$g_{ab}$ is, of course, irrelevant in this context and that the metric
in question therefore certainly does not need to specifically match
the Kerr-Newman metric or any other metric considered in this section.

On the other hand, one must be a little more careful when generalizing
metrics of type $(75)$. In this case, it is favorable to consider
only rescalings $l_{a}\rightarrow\lambda l_{a}$, $k_{a}\rightarrow\lambda^{-1}k_{a}$
and null rotations $l_{a}\rightarrow l_{a},$ $m_{a}\rightarrow m{}_{a}+\xi l_{a},$
$k_{a}\rightarrow k_{a}+\xi\bar{m}_{a}+\bar{\xi}m_{a}+\vert\xi\vert^{2}l_{a}$,
because otherwise the resulting local geometry of spacetime would
no longer be a Kerr-Schild geometry and thus similar problems could
occur as in the case of spacetimes of the generalized Gordon class.
In this way, very general metrics of type $(77)$ can be obtained,
which allow a simple generalization of shock wave geometries of type
$(74)$, which in turn can be constructed from known shock wave geometries
of type $(66)$.

In any case, there are also other useful ways to construct interesting
models for ambient spacetimes $\bar{(\mathcal{M}},\bar{g})$, e.g.
by considering other more complex types of metric deformations. As
a specific example example, superimposed generalized Kerr-Schild deformations
may be mentioned, not least because these types of metric deformations
allow an extension of thin shell formalism (although one must be very
careful with the calculation of curvature expressions). These types
of metric deformations are of the form 
\begin{equation}
\bar{g}_{ab}=g_{ab}+\underset{A=1}{\overset{N}{\sum}}f_{(A)}l_{a}^{(A)}l_{b}^{(A)},
\end{equation}
where each $l_{a}^{(A)}=a^{(A)}l_{a}+b^{(A)}k_{a}+c^{(A)}\bar{m}_{a}+\bar{c}^{(A)}m_{a}$
must meet the conditions $l_{a}^{(A)}l_{(A)}^{a}=0$ and $(l_{(A)}\nabla)l_{(A)}^{a}=0$.
Representatives of this class lead to a whole series of nested Kerr-Schild
spacetimes, i.e.

\begin{align}
\bar{g}_{ab} & =g_{ab}^{(1)}+\underset{A=2}{\overset{N}{\sum}}f_{(A)}l_{a}^{(A)}l_{b}^{(A)}=\\
 & =g_{ab}^{(2)}+\underset{A=3}{\overset{N}{\sum}}f_{(A)}l_{a}^{(A)}l_{b}^{(A)}=...=g_{ab}^{(N-1)}+f_{(N)}l_{a}^{(N)}l_{b}^{(N)},
\end{align}
 where $g_{ab}^{(1)}=g_{ab}+f_{(1)}l_{a}^{(1)}l_{b}^{(1)}$, $g_{ab}^{(2)}=g_{ab}+\underset{A=1}{\overset{2}{\sum}}f_{(A)}l_{a}^{(A)}l_{b}^{(A)}$,
...., $g_{ab}^{(N-1)}=g_{ab}+\underset{A=1}{\overset{N-1}{\sum}}f_{(A)}l_{a}^{(A)}l_{b}^{(A)}$
applies by definition. As highlighted in several works on the subject
\cite{bonning2003physics,marronetti2000solving,matzner1998initial,moreno2002kerr,racz2018simple},
the corresponding metric deformations can be used to provide initial
data for the construction of solutions to Einstein's equations that
characterize multiple black holes in General Relativity. On a non-numerical
level, these types of metric deformations have been used to construct
generalizations of the gravitational shock wave spacetimes previously
considered in this section \cite{huber2019ultpar}. 

However, there are many other classes of both exact and approximate
metric deformations besides the geometric models mentioned above,
which can be used to construct local spacetimes. The deformation of
the background metric can also be applied perturbatively, which can
lead to physically interesting examples. This is not least due to
the fact that the deformation approach allows the geometric transition
of arbitrary spacetimes and thus the gluing of perturbative metrics. 

Because of all these points, the geometric deformation approach represents
a useful extension of the well-established thin shell formalism, which
allows the treatment of a larger class of problems than previously
possible.

\section*{Summary}

In the present work, a specific approach to the construction of local
spacetimes in General Relativity was presented. This approach is based
on the idea of using local deformations of the metric to join spacetimes
with different geometries and physical properties. As it turned out
in this context, the approach presented allows the calculation of
the curvature fields of spacetimes with metrics of low regularity,
such as gravitational shock wave spacetimes, which, from the point
of view of standard gluing techniques, does not seem feasible (or
even possible). Furthermore, it was found that smooth gluings of arbitrary
spacetime pairs can be carried out by using suitable transition functions
and that complex types of ambient metrics and associated spacetimes
can be constructed by transformations that leave the local geometric
structure of spacetime invariant. Not least because the above mentioned
results do not only apply to exact deformations, but also to local
metric perturbations, it can be plausibly concluded from the observations
made in this work that there may be valuable extensions of already
known geometric models of Einstein-Hilbert gravity or more general
gravitational theories that have remained undiscovered so far. It
might therefore be worthwhile to use the geometric deformation approach
in the analysis of complex geometric transitions in spacetime, especially
if these transitions cannot be treated on the basis of the traditional
thin shell formalism.

\bibliographystyle{plain}
\bibliography{0C__Arbeiten_Fertige_Arbeiten__Arxive_lit}

\begin{thebibliography}{10}

\bibitem{aichelburg1971gravitational}
{Peter}~{C} Aichelburg and {Roman}~{Ulrich} {Sexl}.
\newblock On the gravitational field of a massless particle.
\newblock {\em General {Relativity} and {Gravitation}}, 2(4):303--312, 1971.

\bibitem{baccetti2012gordon}
{Valentina} Baccetti, {Prado} {Martin}-{Moruno}, and {Matt} {Visser}.
\newblock Gordon and {Kerr}-{Schild} ans{\"a}tze in massive and bimetric
  gravity.
\newblock {\em Journal of {High} {Energy} {Physics}}, 2012(8):1--19, 2012.

\bibitem{barcelo2001analogue}
{Carlos} Barcelo, {Stefano} {Liberati}, and {Matt} {Visser}.
\newblock Analogue gravity from {Bose}-{Einstein} condensates.
\newblock {\em Classical and {Quantum} {Gravity}}, 18(6):1137, 2001.

\bibitem{barrabes1991thin}
{C} Barrabes and {W} {Israel}.
\newblock Thin shells in general relativity and cosmology: {The} lightlike
  limit.
\newblock {\em Physical {Review} {D}}, 43(4):1129, 1991.

\bibitem{bonning2003physics}
{Erin} Bonning, {Pedro} {Marronetti}, {David} {Neilsen}, and {Richard}
  {Matzner}.
\newblock Physics and initial data for multiple black hole spacetimes.
\newblock {\em Physical {Review} {D}}, 68(4):044019, 2003.

\bibitem{bonnor1970spherically}
{W}{B} Bonnor and {P}{C} {Vaidya}.
\newblock Spherically symmetric radiation of charge in {Einstein}-{Maxwell}
  theory.
\newblock {\em General {Relativity} and {Gravitation}}, 1(2):127--130, 1970.

\bibitem{brill1994gravitational}
{Dieter} Brill and {Geoff} {Hayward}.
\newblock Is the gravitational action additive?
\newblock {\em Physical {Review} {D}}, 50(8):4914, 1994.

\bibitem{brown1993quasilocal}
{J}~{David} Brown and {James}~{W} {York}~{Jr}.
\newblock Quasilocal energy and conserved charges derived from the
  gravitational action.
\newblock {\em Physical {Review} {D}}, 47(4):1407, 1993.

\bibitem{carter1971axisymmetric}
{Brandon} Carter.
\newblock Axisymmetric black hole has only two degrees of freedom.
\newblock {\em Physical {Review} {Letters}}, 26(6):331, 1971.

\bibitem{clarke1987junction}
{C}{J]}{S} Clarke and {Tevian} {Dray}.
\newblock Junction conditions for null hypersurfaces.
\newblock {\em Classical and {Quantum} {Gravity}}, 4(2):265, 1987.

\bibitem{colombeau2000new}
{Jean}~{Fran{\c{c}}ois} Colombeau.
\newblock {\em New generalized functions and multiplication of distributions}.
\newblock Elsevier, 2000.

\bibitem{colombeau2011elementary}
{Jean}~{Fran{\c{c}}ois} Colombeau.
\newblock {\em Elementary introduction to new generalized functions}.
\newblock Elsevier, 2011.

\bibitem{darmois1927equations}
{Georges} Darmois.
\newblock Les {\'e}quations de la gravitation einsteinienne.
\newblock {\em M{\'e}morial des sciences math{\'e}matiques}, 25:1--48, 1927.

\bibitem{de2014massive}
{Claudia} de~{Rham}.
\newblock {Massive} {Gravity}.
\newblock {\em Living {Rev.} {Rel}}, 17(7):1401--4173, 2014.

\bibitem{dray1985gravitational}
{Tevian} Dray and {Gerard't} {Hooft}.
\newblock The gravitational shock wave of a massless particle.
\newblock {\em Nuclear physics {B}}, 253:173--188, 1985.

\bibitem{fagnocchi2010relativistic}
{Serena} Fagnocchi, {Stefano} {Finazzi}, {Stefano} {Liberati}, {Marton}
  {Kormos}, and {Andrea} {Trombettoni}.
\newblock Relativistic {Bose}-{Einstein} condensates: a new system for analogue
  models of gravity.
\newblock {\em New {Journal} of {Physics}}, 12(9):095012, 2010.

\bibitem{fierz1939relativistic}
{Markus} Fierz and {Wolfgang} {Pauli}.
\newblock On relativistic wave equations for particles of arbitrary spin in an
  electromagnetic field.
\newblock {\em Proceedings of the {Royal} {Society} of {London}. {Series} {A},
  {Mathematical} and {Physical} {Sciences}}, pages 211--232, 1939.

\bibitem{gordon1923lichtfortpflanzung}
{Walter} Gordon.
\newblock Zur {Lichtfortpflanzung} nach der {Relativit{\"a}tstheorie}.
\newblock {\em Annalen der {Physik}}, 377(22):421--456, 1923.

\bibitem{grosser2001geometric}
Grosser, {Kunzinger}, {Oberguggenberger}, and {Steinbauer}.
\newblock {\em Geometric theory of generalized functions with applications to
  general relativity}, volume 537.
\newblock Springer {Science} \& {Business} {Media}, 2001.

\bibitem{hassan2012bimetric}
{S}~{Fawad} Hassan and {Rachel}~{A} {Rosen}.
\newblock Bimetric gravity from ghost-free massive gravity.
\newblock {\em Journal of {High} {Energy} {Physics}}, 2012(2):1--12, 2012.

\bibitem{hawking1973large}
{Stephen}~{W} Hawking and {George} {Francis}~{Rayner} {Ellis}.
\newblock {\em The large scale structure of space-time}, volume~1.
\newblock Cambridge university press, 1973.

\bibitem{hayward1993gravitational}
{Geoff} Hayward.
\newblock Gravitational action for spacetimes with nonsmooth boundaries.
\newblock {\em Physical {Review} {D}}, 47(8):3275, 1993.

\bibitem{hossenfelder2017analogue}
{Sabine} Hossenfelder and {Tobias} {Zingg}.
\newblock Analogue gravity models from conformal rescaling.
\newblock {\em Classical and quantum gravity}, 34(16):165004, 2017.

\bibitem{huber2020distributional}
{Albert} Huber.
\newblock Distributional metrics and the action principle of
  {Einstein}-{Hilbert} gravity.
\newblock {\em Classical and {Quantum} {Gravity}}, 37(8).

\bibitem{huber2019ultpar}
{Albert} Huber.
\newblock The gravitational {Field} of a massless {Particle} on the {Horizon}
  of a stationary {Black} {Hole}.
\newblock {\em arXiv:1911.02299}.

\bibitem{isham1971nonlinear}
{Chris}~{J} Isham, {Abdus} {Salam}, and {J} {Strathdee}.
\newblock Nonlinear realizations of space-time symmetries. {Scalar} and tensor
  gravity.
\newblock {\em Annals of {Physics}}, 62(1):98--119, 1971.

\bibitem{israel1966singular}
{Werner} Israel.
\newblock Singular hypersurfaces and thin shells in general relativity.
\newblock {\em Il {Nuovo} {Cimento} {B} {Series} 10}, 44(1):1--14, 1966.

\bibitem{koh1992distributions}
EL~Koh and Li~Chen Kuan.
\newblock On the distributions $\delta$k and ($\delta$') k.
\newblock {\em Mathematische Nachrichten}, 157(1):243--248, 1992.

\bibitem{li2014defining}
Chenkuan Li and Changpin Li.
\newblock On defining the distributions $\delta$k and ($\delta$′) k by
  fractional derivatives.
\newblock {\em Applied Mathematics and Computation}, 246:502--513, 2014.

\bibitem{li2007review}
CK~Li.
\newblock A review on the products of distributions.
\newblock In {\em Mathematical methods in engineering}, pages 71--96. Springer,
  2007.

\bibitem{lousto1989ultrarelativistic}
{C}{O} Lousto and {N} {S{\'a}nchez}.
\newblock The ultrarelativistic limit of the {Kerr}-{Newman} geometry and
  particle scattering at the {Planck} scale.
\newblock {\em Physics {Letters} {B}}, 232(4):462--466, 1989.

\bibitem{lousto1990curved}
{C}{O} Lousto and {N} {S{\'a}nchez}.
\newblock The curved shock wave space-time of ultrarelativistic charged
  particles and their scattering.
\newblock {\em International {Journal} of {Modern} {Physics} {A}},
  5(05):915--938, 1990.

\bibitem{marronetti2000solving}
{Pedro} Marronetti and {Richard}~{A} {Matzner}.
\newblock Solving the initial value problem of two black holes.
\newblock {\em Physical {Review} {Letters}}, 85(26):5500, 2000.

\bibitem{mars1993geometry}
{Marc} Mars and {Jose}~{M}{M} {Senovilla}.
\newblock Geometry of general hypersurfaces in spacetime: junction conditions.
\newblock {\em Classical and {Quantum} {Gravity}}, 10(9):1865, 1993.

\bibitem{matzner1998initial}
Richard~A Matzner, Mijan~F Huq, and Deirdre Shoemaker.
\newblock Initial data and coordinates for multiple black hole systems.
\newblock {\em Physical Review D}, 59(2):024015, 1998.

\bibitem{moreno2002kerr}
{Claudia} Moreno, {Dar{\'\i}o} {N{\'u}{\~n}ez}, and { Olivier} {Sarbach}.
\newblock Kerr-{Schild}-type initial data for black holes with angular momenta.
\newblock {\em Classical and {Quantum} {Gravity}}, 19(23):6059, 2002.

\bibitem{michael1992multiplication}
Michael Oberguggenberger.
\newblock {\em Multiplication of distributions and applications to partial
  differential equations}.
\newblock Number 259. Longman Scientific \& Technical, 1992.

\bibitem{penrose1972general}
R~Penrose.
\newblock General relativity, papers in honour of jl synge.
\newblock {\em Clarendon, Oxford}, page 101, 1972.

\bibitem{racz2018simple}
{Istv{\'a}n} R{\'a}cz.
\newblock A simple method of constructing binary black hole initial data.
\newblock {\em Astronomy {Reports}}, 62(12):953--958, 2018.

\bibitem{reina2016junction}
{Borja} Reina, {Jos{\'e}}~{M}{M} {Senovilla}, and {Ra{\"u}l} {Vera}.
\newblock Junction conditions in quadratic gravity: thin shells and double
  layers.
\newblock {\em Classical and Quantum Gravity}, 33(10):105008, 2016.

\bibitem{robinson1975uniqueness}
{David}~{C} Robinson.
\newblock Uniqueness of the {Kerr} black hole.
\newblock {\em Physical {Review} {Letters}}, 34(14):905, 1975.

\bibitem{robinson1974classification}
{D}{C} Robinson.
\newblock Classification of black holes with electromagnetic fields.
\newblock {\em Physical {Review} {D}}, 10(2):458, 1974.

\bibitem{senovilla2015double}
{Jos{\'e}}~MM Senovilla.
\newblock Double layers in gravity theories.
\newblock In {\em Journal of {Physics}: {Conference} {Series}}, volume 600,
  page 012004. I{O}{P} {Publishing}, 2015.

\bibitem{senovilla2018equations}
{Jos{\'e}}~{M}{M} Senovilla.
\newblock Equations for general shells.
\newblock {\em Journal of {High} {Energy} {Physics}}, 2018(11):134, 2018.

\bibitem{sfetsos1995gravitational}
{Konstadinos} Sfetsos.
\newblock On gravitational shock waves in curved spacetimes.
\newblock {\em Nuclear {Physics} {B}}, 436(3):721--745, 1995.

\bibitem{taub1980space}
{A}{H} Taub.
\newblock Space-times with distribution valued curvature tensors.
\newblock {\em Journal of Mathematical Physics}, 21(6):1423--1431, 1980.

\bibitem{taub1981generalised}
AH~Taub.
\newblock Generalised kerr-schild space-times.
\newblock {\em Annals of Physics}, 134(2):326--372, 1981.

\bibitem{unruh1981experimental}
{William}~{George} Unruh.
\newblock Experimental black-hole evaporation?
\newblock {\em Physical {Review} {Letters}}, 46(21):1351, 1981.

\bibitem{visser2010acoustic}
{Matt} Visser and {Carmen} {Molina}-{Par{\'\i}s}.
\newblock Acoustic geometry for general relativistic barotropic irrotational
  fluid flow.
\newblock {\em New {Journal} of {Physics}}, 12(9):095014, 2010.

\end{thebibliography}

\end{document}